\documentclass[12pt,a4paper,twoside,fleqn,epsf]{article}
\usepackage{latexsym}
\usepackage{emlines}
\usepackage{epsf}
\usepackage{feynmf}

\def\be{\begin{equation}}
\def\ee{\end{equation}}
\def\bea{\begin{eqnarray}}
\def\eea{\end{eqnarray}}
\def\Litwo{{\mbox{\rm Li}_2}}
\def\indone{\;\;\;\;\;\;\;\;}
\def\indtwo{\;\;\;\;\;\;\;\;\;\;\;}
\def\indthree{\;\;\;\;\;\;\;\;\;\;\;\;\;\;\;\;\;\;\;\;\;\;\;\;}

\def\OLL{{\cal O}_{LL}}
\def\ol{\overline}
\begin{document}

\thispagestyle{empty}

\vspace{0.5cm}

\begin{center}
{\Large Next--to--Leading Order QCD corrections}\\
{\Large to $B\overline B$--mixing and $\epsilon_K$ within the MSSM}\\[5mm]
F. Krauss, G. Soff
\end{center}
\vspace{5mm}

\centerline
{Institut f$\ddot{\rm u}$r Theoretische Physik}
\centerline
{Technische Universit$\ddot{\rm a}$t Dresden}
\centerline
{Mommsenstr.13, D-01062 Dresden, Germany}

\vspace{0.5cm}

\begin{center}
\parbox{15cm}{
{\centerline\Large Abstract}

\vspace{0.5cm}

{\footnotesize
We present a calculation of the QCD correction factors $\eta_{2B}$ and 
$\eta_{2K}$ up to Next--to--Leading Order within the MSSM. 
We took into account the region of low $\tan\!\beta$ for the 
Higgs-- and chargino sector while neglecting the effect of gluinos and 
neutralinos.
}}

\vspace{0.5cm}
\noindent
Key--words: Mixing; MSSM; CP-violation; CKM-matrix 
\end{center}

\section{Introduction}

During the last decade, Supersymmetry (SUSY) has been one of the most
popular candidates for physics beyond the Standard Model (SM), see, e.g.,
the reviews \cite{Hab85}-\cite{Hab93}. 
Among all other ideas SUSY seemingly has the greatest potential to
allow for the construction of a Grand Unified Theory. 
Even more, a low--energy version of SUSY would cancel the quadratic 
divergencies emerging within the Higgs--sector and cure the most nagging 
shortcomings of the Standard Model \cite{Mai80}-\cite{Daw97}. 
However, so far there is no experimental evidence for any of the new 
particles predicted by the various supersymmetric models up to scales of 
the order of the $W$-- and $Z$--mass.
So before the advent of the next generation of new colliders
one probably has to rely on indirect probes to test the idea
of a low--energy SUSY. 

In this respect the most promising candidates are processes involving Flavour 
Changing Neutral Currents (FCNC), especially rare decays like
$b\to s\gamma$ and oscillations in the system of neutral mesons, namely 
$B\ol{B}-$ and $K\ol{K}-$mixing \cite{Ber91}-\cite{Hew98}. In this paper we will 
focus on the latter class of processes and consider contributions to
the mixing phenomena mediated by the chargino--squark and the
charged Higgs--up--type quark sector of
the Minimal Supersymmetric Standard Model (MSSM) taking into account 
gluonic Next--to--Leading Order QCD--corrections (NLO). We present
results for the mass--splitting $\Delta M_B$ within the corresponding
neutral $B$--system and for the parameter $\epsilon_K$ describing 
CP--violation in the $K$--system.

\begin{figure}\begin{center}
\,\mbox{\epsfxsize=14cm\epsffile{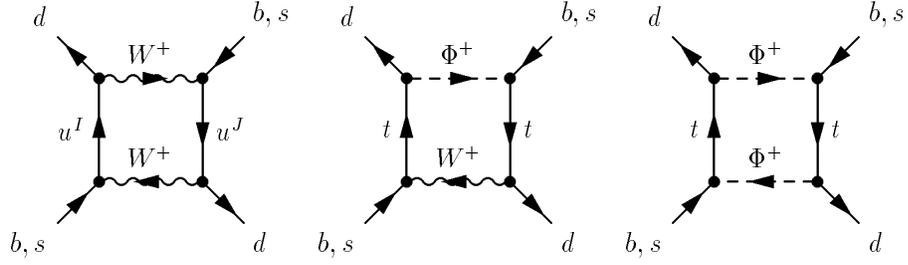}}
\parbox[t]{13.5cm}{\caption{\label{FigSMLO}
\footnotesize Diagrams contributing to 
$B\ol{B}$--mixing in LO within the SM. The would--be Goldstone bosons $\Phi$ 
emerging when using the Feynman--t'Hooft gauge are displayed explicitly,
crossed diagrams are missing.}}
\end{center}\end{figure}

Within the framework of the SM the corresponding $\Delta F = 2$ 
processes are governed
by box--diagrams with internal up--type quarks and $W$--bosons and are 
suppressed by either small quark masses ($u$ and $c$) when compared to
the $W$--mass or by small mixings between the third and the first
two generations. The corresponding Leading Order diagrams of the SM are 
depicted in Fig.\,\ref{FigSMLO}. The result for the SM up to NLO QCD 
corrections can be found in \cite{Bur90}-\cite{Urb98}.

\begin{figure}\begin{center}
\,\mbox{\epsfxsize=10cm\epsffile{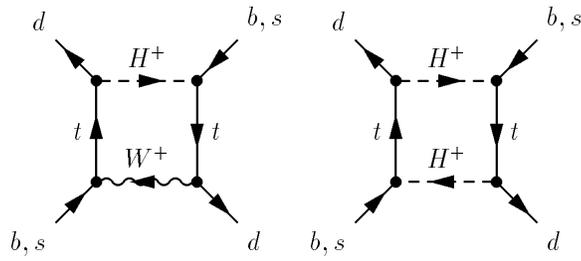}}
\parbox[t]{9cm}{\caption{\label{FigTHDMLO}
\footnotesize Diagrams contributing to 
$B\ol{B}$--mixing in LO within the THDM. The would--be Goldstone bosons 
$\Phi$ are subsumed in the labels $W$, crossed diagrams are missing.}}
\end{center}\end{figure}

\begin{figure}\begin{center}
\,\mbox{\epsfxsize=14cm\epsffile{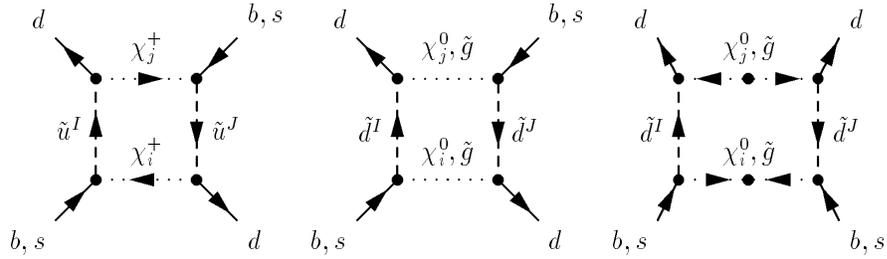}}
\parbox[t]{13.5cm}{\caption{\label{FigSUSYLO}
\footnotesize Diagrams contributing to 
$B\ol{B}$--mixing in LO within the MSSM. Only the boxes involving 
supersymmetric particles are displayed. Note that the neutralinos
and gluinos are Majorana--particles. This fact produces graphs like
the last one displayed where we have shown explicitly the effect
of the Majorana's character by adding the dots and the direction of 
the flow with arrows. This feature results in scalar operators.}}
\end{center}\end{figure}

Within the MSSM -- the supersymmetric extension of the SM with
minimal field content, the same gauge group and soft symmetry breaking --
there are different sources of flavour--changing processes.
Nevertheless, all of them enter the stage of $\Delta F = 2$ transitions 
via box--diagrams.
First of all there are loops with internal up--type quarks and charged 
Higgs--Bosons (see Fig.\,\ref{FigTHDMLO}). Taken alone, this corresponds 
to the Two Higgs Doublet Model (THDM), another popular extension of the SM.
Results up to Next--to--Leading Order have been presented in \cite{Urb98}. 
The second source are loops involving charginos and up--type squarks
(see Fig.\,\ref{FigSUSYLO}). 
Their effect is investigated in this article up to NLO of 
gluonic QCD corrections. 
The last source of the $\Delta F = 2$ processes within the MSSM are 
off--diagonal elements within the squark--mass matrices, entering via boxes 
containing down--type squarks and gluinos or neutralinos 
(see again Fig.\,\ref{FigSUSYLO}). 
In the context of this paper we do
not take into account their effect. Additionally we do not consider
supersymmetric QCD--corrections mediated by gluinos although their effect
might be sizable even in the limit of heavy gluinos \cite{Ciu98}. The 
incorporation of these contributions would require additional extensive
calculations which is beyond the scope of the current paper. 
In Leading Order the effect of the various MSSM--contributions on the
mixing phenomena has been examined in \cite{Gab96}-\cite{Bag97}. 
There it is shown, that $K\ol{K}$--mixing strongly limits the size of some 
off--diagonal elements within the down--type squark matrix.
Additionally, within the context of the mSUGRA model 
\cite{Cha82}-\cite{Nat83}, 
LO--results have been presented in \cite{GNO95}. In this specific version
of a supersymmetric extension of the SM, the soft supersymmetry--breaking
terms are governed by just 5 parameters, reducing the vast number of 
parameters occuring within the MSSM considerably. The authors of
\cite{GNO95} claim, that the most sizable contributions to $\Delta M_B$ and 
$\epsilon_K$ of the mSUGRA model in addition to the SM stem from the 
chargino--squark and the charged Higgs--up--type quark boxes when taking
into account constraints of the parameter space as given by various other
processes. These findings partly justify our focus on the charged Higgs-- and
the chargino--sector as primary sources of FCNC--processes.

It should be stressed here, that a priori the ratio of the vacuum expectation
values of the two Higgs doublets, $\tan\!\beta$, is a free parameter.
Overall fits to available experimental data constrain this parameter to
two regions, $\tan\!\beta \approx 30-50$ and $\tan\!\beta\approx 1-2$,
and it seems that the latter one is the more favourable one
\cite{Boe97,Boe97a,Erl98}. Here we will concentrate on this region of 
low $\tan\!\beta$, which yields the major contributions to the
mixing phenomena as stated by \cite{GNO95}.

The paper is organized as follows. After we redisplay the necessary parts 
of the MSSM--Feynman rules we shortly review the basic formalism in LO. 
In section 3 we discuss some features of the NLO
calculations with special focus on the matching--procedure. In section
4 we provide the results in NLO for the different observables and scan to some
extent the parameter space of the MSSM. We close with some concluding
remarks.

\section{Notation and basic formalism}

\subsection{Notation and Feynman--Rules}

Throughout this paper we use the notation of \cite{Ros90}. Here it should 
be sufficient to list the relevant Feynman--rules and define
the quantities involved. For the gauge bosons we employ the Feynman--t'Hooft 
gauge $\xi = 1$, the propagators can be found in the Appendix.
The relevant vertices are depicted in \ref{Feyns}. There, the capital 
letters $I,J,K$ denote the quark generations, the letters $i,j$ 
label the squark--, Higgs-- or chargino--fields and the $a,b,c$ are SU(3)
indices in the appropriate representation. The matrices $Z^\pm$ and 
$Z_{U,D}$ diagonalize the mass matrices $X$ of the charginos and the up-- 
and down--type squarks respectively,
\be
(Z^-)^T X_\chi Z^+ = \mbox{\rm diag}(m_{\chi_{1,2}})\;\;, 
Z^\dagger_{U,D} X_{U,D} Z_{U,D} = \mbox{\rm diag} 
(m_{\tilde{u_I},\tilde{d_I}}),
\ee
Throughout this paper we will assume a specific form of the matrix
$Z_U^\pm$ responsible for the mixing of the different squarks, namely
all off--diagonal elements zero besides the ones associated with
the mixing of the stops, $\tilde{u}^{3,6}$. Additionally, we assume
the up--type squarks of the first two generations degenerate in mass.
In other words, we assume, that the squark mass--matrix preserves flavour and 
mixes only the stops of different chiralities. This special form of the 
$Z$--matrices used in this paper is given in Eq.\,(\ref{MSSMmatrices}).
It should be stressed here, that the last two assumptions concerning the
up--type squark sector are mainly for the sake of a compact presentation but
still somewhat unfounded. However, a generalization of our results to a more 
complicated structure of the up--type squark mass--matrix is quite 
straight--forward. 
It should be noted, that the four--squark vertex containing two up--type
and two down--type squarks is not of relevance when considering
${\cal O}(\alpha_s)$--corrections.  

$C^{IJ}$ denotes the CKM--element between generations $I$ and $J$ and 
$\tan\!\beta = v_2/v_1$ is the ratio of the vacuum expectation values of the 
corresponding Higgs--doublets. We have explicitly written down the Yukawa--type
couplings in terms of the quark masses involved. Simple inspection shows, 
that for the region of low $\tan\!\beta$ we can neglect the couplings
proportional to the down--type masses with respect to the top--mass.

We use the short--hand notation 
\be
P_{L,R} = \frac{1\mp\gamma_5}{2}\,,\gamma^{\mu L} = \gamma^\mu P_L
\ee
for the left-- and right--handed projectors and their combinations
with gamma--matrices.
 
\subsection{Leading Order}

We want to summarize now briefly the ingredients used to obtain Leading Order 
results. 
In case of the $B\ol{B}$--mixing within the Standard Model this is quite
transparent. Neglecting all masses besides the $W$-- and the top--mass
one can use the GIM--mechanism to account for the two lighter quark--types
running in the boxes and calculate the remaining box--diagrams. This
is simplified even more by using the Feynman--t'Hooft gauge for the $W$,
leaving us with the physical $W$--boson and the would--be Goldstone--boson
$\Phi$, both with identical masses. Results for boxes involving additional
Higgs--bosons and top--quarks can be obtained in the same manner.
Due to the fact that the supersymmetric partners of the SM--particles
differ in Spin by $1/2$ one has to perform a Fierz--transformation 
for diagrams with squark-- and chargino--lines to write down the emerging
operator in terms of colour--singlet currents. As stated before, we assume, 
that the squarks of the first two generations ($\tilde u_i\,,i=1,2,4,5$) 
are degenerate in mass and do not mix considerably. This allows 
us to use the GIM--mechanism within the squark sector to 
tighten our final results. Generalization to non--degenerate squarks is 
straightforward.

\begin{figure}\begin{center}
\,\mbox{\epsfxsize=14cm\epsffile{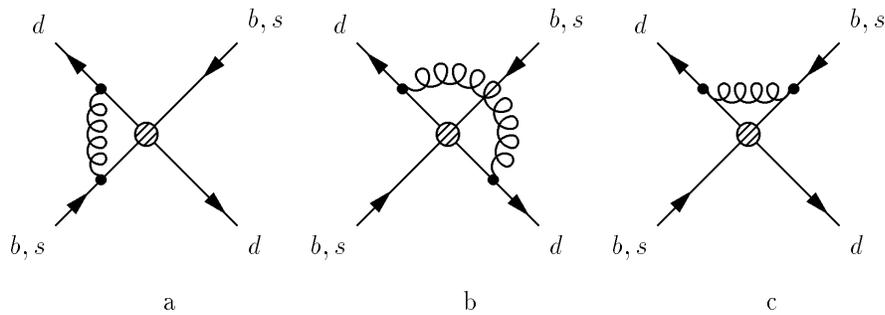}}
\parbox[t]{14cm}{\caption{\label{Figeta2}
\footnotesize Diagrams contributing to 
the factor $\eta_2$ for the QCD--corrections in LO.}}
\end{center}\end{figure}

The procedure of integrating out the heavy degrees of freedom leaves us 
with one operator and its
Wilson--coefficient at the matching scale ${\mu_0} = O(M_W^2)$ and yields
the starting condition for the renormalization group equations.
The connection between the starting scale and the bosonic scale 
$\mu = O(m_B^2)$ is given in LO by the diagrams of Fig.\,\ref{Figeta2} 
and a corresponding factor $\eta_{2}$. Putting everything
together, the effective Hamiltonian for $\Delta B = 2$--transitions at the 
bosonic scale is given by
\be
\label{heffB}
H_{\rm{eff}} \;= \frac14
    \,{G^2_F \over \pi^2}\; m_W^2 \;{( V_{td}V^\ast_{tb})}^2 
    \;\eta_{2}\; S \;\OLL(\mu)\;,
\ee
where
\be
\OLL =\left(\bar{d}_i \gamma_{\mu L} b_i \right) \;
      \left(\bar{d}_j \gamma^{\mu L} b_j \right)\;.
\ee
We have written out the colour indices for the quark fields explicitly.
The coefficient $S$ is a sum of the well--known Inami--Lim functions 
\cite{IL81} for the different internal particles,
\bea\label{S1}
S &=& S(x_W,x_H) + \tilde S(\{x_i,y_a\})
\nonumber\\ 
S(x_W,x_H) &=& S_{WW}(x_W)+2\,S_{WH}(x_W,x_H)+S_{HH}(x_H)\;.
\nonumber\\
\tilde S(\{x_i,y_a\}) &=&
            \sum\limits_{i,j=1}^2
            \sum\limits_{a,b=1}^6 \tilde K_{ij,ab}
                 \tilde S(x_i,x_j,y_a,y_b)
\eea
Throughout this paper we use the following abbreviations
\be
x_{W,H} = \frac{m_t^2}{M_{W,H}^2}\;,\;
x_{d,s,c,b} = \frac{m_{d,s,c,b}^2}{M_W^2}\;,\;
y_a = \frac{\tilde m_{\tilde q_a}^2}{M_W^2}\;,\;
x_i = \frac{\tilde m_{\tilde \chi_i}^2}{M_W^2}
\ee
for the ratios of masses entering the Inami--Lim functions. 

The $\tilde K$ in Eq.\,(\ref{S1}) account for the squark--quark--chargino 
couplings as given in the Feynman--rules above. They read
\be
\tilde K_{ij,ab} = 
\prod\limits_{c=\{a,b\}}\prod\limits_{k=\{i,j\}}
\left(Z_U^{Jc}Z_{1k}^-
\frac{m_u^J}{\sqrt{2}M_W\sin\!\beta}Z_U^{(J+3)c}Z^{-}_{2k}\right)\,.
\ee

For the $K$--system the situation is slightly different. Because of the
much lower mass of the $K$--meson, the charm--mass can not be neglected 
any more, and one has to account for this fact by integrating out the heavy 
degrees of freedom in consecutive steps \cite{GW80}. So we end with a 
modified effective Hamiltonian for $\Delta S = 2$--transitions at the bosonic
scale, namely
\bea
\label{heffK}
H_{\rm{eff}} &=& \frac14
    \,{G^2_F \over \pi^2}\; m_W^2 \;
    \left[{( V_{cd}V^\ast_{cs})}^2 \eta_1\,S(x_c) + 
          {( V_{td}V^\ast_{ts})}^2 \eta_2\,S  \right.
\nonumber\\&&\indtwo\indone \left.
     +(V_{cd}V^\ast_{cs})( V_{td}V^\ast_{ts}) \eta_3\,S(x_c,x_W) 
     \right]\;\OLL(\mu)\;,
\eea
where the contribution of the heavy particles alone, $S$ is given by 
Eq.\,(\ref{S1}) and the contributions involving a $c$--quark are
contained in $S(x_c)$ and $S(x_c,x_W)$ with corresponding QCD--correction
factors $\eta_{1,3}$. Of course the operator $\OLL$ changes accordingly,
the $b$--quarks are replaced by strange quark fields.

The factors $\eta_{2\{B,K\}}$ in LO are given by \cite{GW80}
\bea\label{etaLO}
\eta_{2B,LO} &=& \alpha_s(M_W)^{\gamma^{(0)}/(2\beta^{(0)}_5)}\;,\;
\nonumber\\
\eta_{2K,LO} &=& \alpha_s(m_c)^{\gamma^{(0)}/(2\beta^{(0)}_3)}
            \left(\frac{\alpha_s(m_b)}
                  {\alpha_s(m_c)}\right)^{\gamma^{(0)}/(2\beta^{(0)}_4)}
            \left(\frac{\alpha_s(M_W)}
                  {\alpha_s(m_b)}\right)^{\gamma^{(0)}/(2\beta^{(0)}_5)}\;,
\eea
where we have explicitly chosen the matching scale ${\mu_0}=M_W$.
$\gamma^{(0)}$ and $\beta^{(0)}_{n_f}$ are the coefficents of the
anomalous dimension of the operator $\OLL$ and the QCD $\beta$--function in 
LO, the latter one is labelled by the number of active flavours, $n_f$. 
They are given by
\be\label{bgLO}
\gamma^{(0)} = 6\frac{N_c-1}{N_c}\;,\;
\beta^{(0)}_{n_f} = \frac{11 N_c - 2n_f}{3}\;.
\ee
$N_c = 3$ denotes the number of colours. We will see in the following, that 
we are able to plug the NLO QCD--corrections within the MSSM into the 
corresponding factor $\eta_2$, see Eqs.\,(\ref{heffB},\ref{heffK}). The other
QCD--factors $\eta_1$ and $\eta_3$ remain unaltered. Because they are 
even in LO quite complicated functions of $\alpha_s$, we just quote the 
numerical results for the NLO expressions we will use later,
\be
\eta_1 \approx 1.38\;,\;\eta_3 \approx 0.47
\ee
as given in \cite{Buc96,Her95}.

The last task left is to extract the physically observable quantities
from the effective Hamiltonians. In the case of the mass difference
$\Delta M_B$ in the $B$--system and $\epsilon_K$ one has to sandwich
the Hamiltonian in between two mesonic states. This amounts to the 
use of the relation
\be
\langle\ol B|\OLL(\mu)|B\rangle = \frac23 B_B(\mu) f_B^2 m_B\,,
\ee
where the so--called bag--parameter is given scale--independently by
\be\label{Bagdef}
B_B = B_B(\mu)\alpha_s(\mu)^{g^{(0)}/(2\beta^{(0)}_{n_f})}
\left[1-\frac{\alpha_s(\mu)}{4\pi}\,Z_{n_f}\right]\,.
\ee
Similar expressions hold for the $K$--system as well. In LO the last
term of equation (\ref{Bagdef}) can be omitted, the term $Z_{n_f}$ 
appearing there will be defined in the next section.

Unfortunately the evaluation of $B_K$ and $B_B f_B^2$ respectively
by lattice calculations yields large uncertainties \cite{Buc96},
\be
B_K = 0.75\pm 0.15\;,\;\sqrt{B_B} f_B^2 = 0.2\pm 0.04\,\mbox{\rm  GeV}\,,
\ee
and it is fair enough to state that this fact spoils the proper
determination of the CKM--element $V_{td}$ by the processes considered
here. 

\section{Explicit QCD--corrections, matching and running}

\subsection{Explicit QCD--corrections}

\begin{figure}\begin{center}
\,\mbox{\epsfxsize=14cm\epsffile{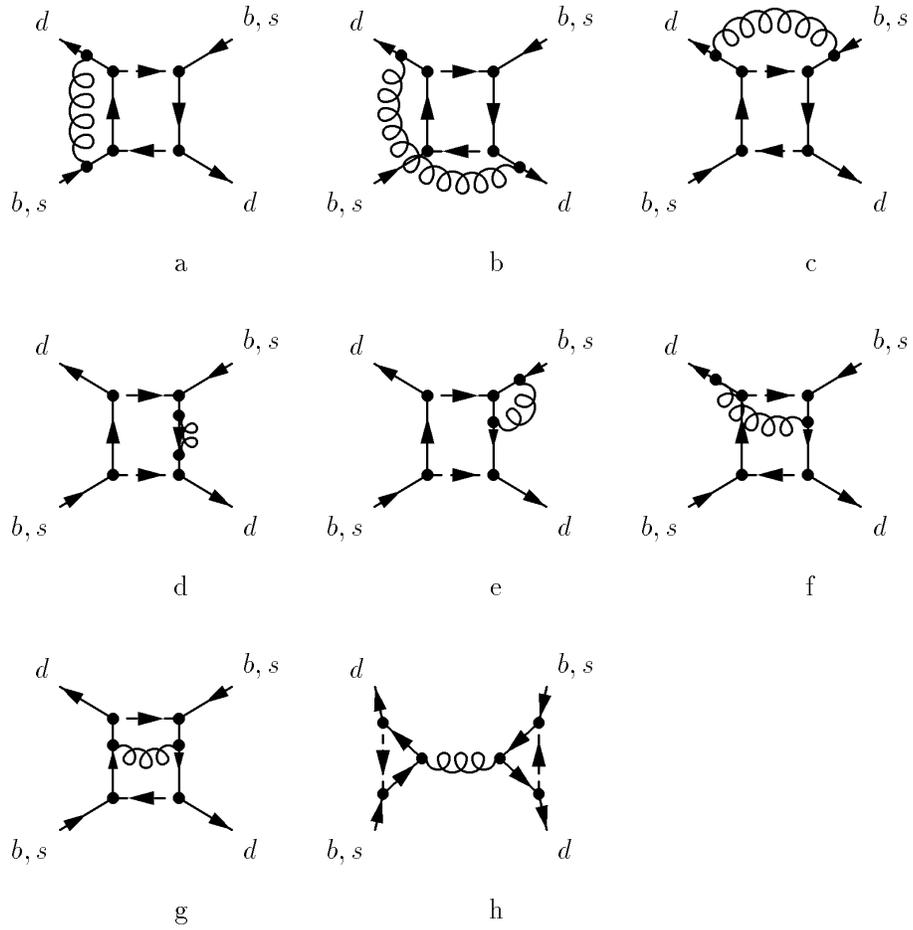}}
\parbox[t]{14cm}{\caption{\label{FigTHDMNLO}
\footnotesize Diagrams responsible for explicit
QCD--corrections contained in $\eta_2$ at NLO within the SM and the THDM.}}
\end{center}\end{figure}

\begin{figure}\begin{center}
\,\mbox{\epsfxsize=14cm\epsffile{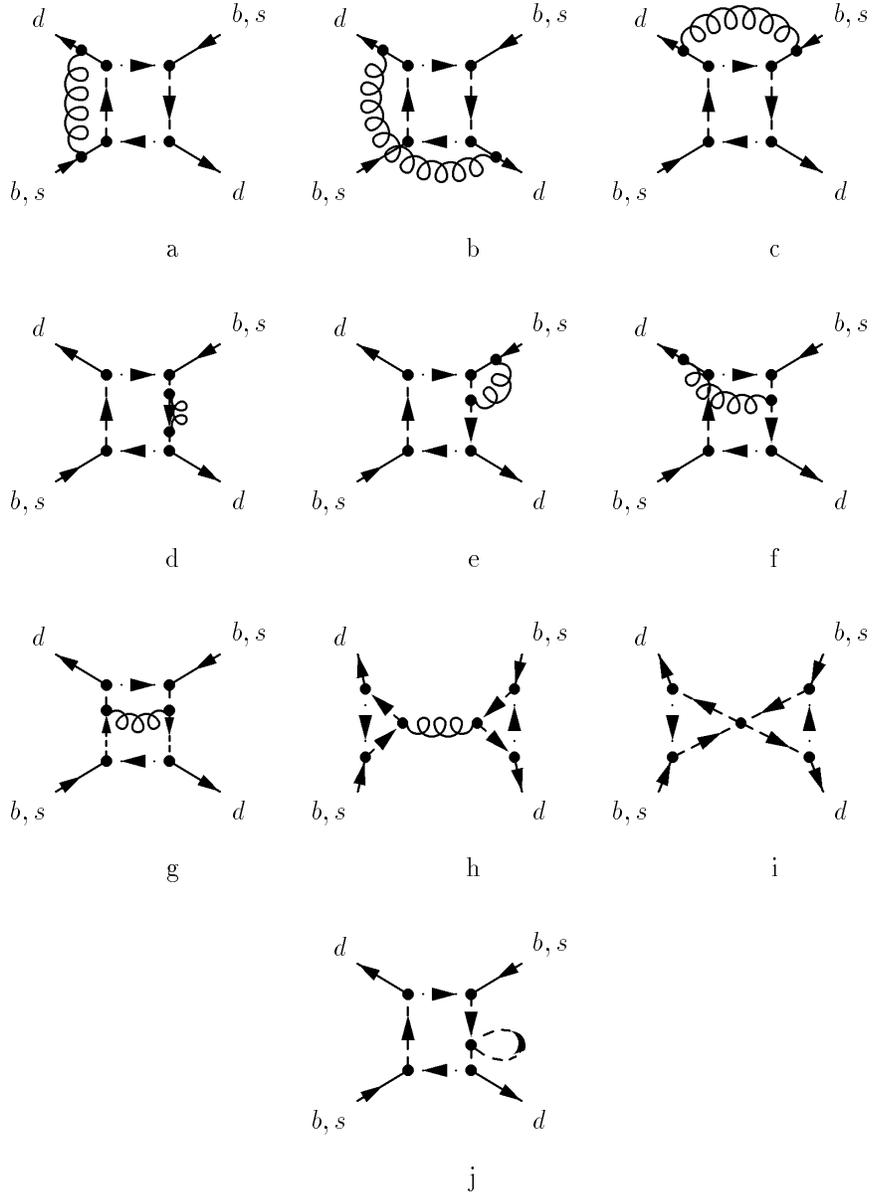}}
\parbox[t]{14cm}{\caption{\label{FigSUSYNLO}
\footnotesize Diagrams responsible for explicit
QCD--corrections contained in $\eta_2$ at NLO within the chargino--squark 
sector of the MSSM.}}
\end{center}\end{figure}

We will now proceed by presenting the explicit perturbative QCD corrections
up to $O(\alpha_s)$. They are obtained by evaluating the diagrams in Figs.\,
\ref{FigTHDMNLO} and \ref{FigSUSYNLO}. For the diagrams containing quarks 
and bosons we performed the calculation in an arbitrary covariant 
$\xi$--gauge for the gluon and -- as stated before -- the Feynman--'tHooft 
gauge for the $W$--boson. For diagrams involving supersymmetric particles we
chose explicitly $\xi = 1$ for the gluon. Again, we perform a 
Fierz--transformation for the latter diagrams to write down the operators 
in terms of colour--singlett currents.

As one easily notices, diagrams (b,c,f,g,h) have the octet--structure
$T_a\otimes T^a$ whereas the diagrams (a,d,e,j) have the singlet--structure
${\bf 1}\otimes {\bf 1}$. The double penguin diagrams (h) and the 
diagram (i) involving the four--squark coupling in a similar fashion do not 
contribute for vanishing external momenta. 

Furthermore, we have to face ultraviolet as well as infrared divergencies. 
The first ones stem from the diagrams (d,e,j) . 
In this case we employ dimensional 
regularization within the so--called Naive Dimensional 
Regularization--scheme (NDR--scheme) \cite{Bur90a,Her95} and we renormalize 
using the $\ol{MS}$--scheme \cite{Bur90a,Bar78}. Note, that for the 
supersymmetric particles we have to encounter additionally tadpole--like
contribution (i) caused by the four--squark vertex. The infrared divergencies
stem from the diagrams (a,b,c) . To handle them, we keep the masses of the 
external quarks whenever necessary. We will see, that this particular choice 
does not affect the final result for the Wilson--coefficient after performing 
the matching. Of course we could have used dimensional regularization as well,
but the method chosen allows to compare the calculation presented here
step by step with the ones presented earlier \cite{Bur90,Urb98}.

In the following we will only consider the contribution to $\eta_2$,
the parts of (\ref{heffK}) involving the charm quark will not be displayed
in this section. This reduces (\ref{heffK}) to (\ref{heffB}), and we will
perform the necessary steps to gain $\eta_{2B}$ at NLO in this section
before we just generalize on $\eta_{2K}$ at the end.

So the $O(\alpha_s)$--corrections to the Hamiltonian of (\ref{heffB}) 
show the structure
\be
\Delta H_{\rm eff} = \frac{G_F^2}{4\pi^2}M_W^2\frac{\alpha_s}{4\pi}
(V_{td}V_{tb}^\ast)^2\,U\,,
\ee
with $U$ given by
\be
U = \sum_k\left({\bf 1}\otimes{\bf 1}\,C_F\phi_k^{(1)}
                      +T_a\otimes T^a\,\phi_k^{(8)}\right){\cal O}_k\,,
\ee
where a sum over $k=1,2,3,LL$ is performed. The operators ${\cal O}_k$ read
\bea
\OLL &=&\left(\bar{d}_i \gamma_{\mu L}b_i \right) 
        \left(\bar{d}_j \gamma^{\mu L}b_j \right),\nonumber\\
{\cal O}_1 &=&\left(\bar{d}_i P_Lb_i \right) 
        \left(\bar{d}_j P_Lb_j \right)-
        \left(\bar{d}_i \sigma_{\mu\nu L}b_i \right) 
        \left(\bar{d}_j \sigma^{\mu\nu L}b_j \right) +
         \{L \leftrightarrow R\}\,,\nonumber\\
{\cal O}_2 &=&\left(\bar{d}_i P_Rb_i \right) 
        \left(\bar{d}_j P_Lb_j \right) +
         \{L \leftrightarrow R\}\,,\nonumber\\
{\cal O}_2 &=&\left(\bar{d}_i \gamma_{\mu R}b_i \right) 
        \left(\bar{d}_j \gamma^{\mu L}b_j \right) +
         \{L \leftrightarrow R\}\,.
\eea
The operators ${\cal O}_{1,2,3}$ stem from the infrared divergent diagrams
b,c and a, respectively. All of them are written in a Fierz--invariant
fashion, allowing us to perform the Fierz--transformation for the
supersymmetric contributions.

The colour factor $C_F$ along with $\tilde{C}_A$ to be used later is given by
\be
C_F = \frac{N_c^2-1}{2N_c}\;,\;
\tilde{C}_A = \frac{N_c-1}{2N_c}\,.
\ee
In analogy to Eq.\,(\ref{S1}) the functions $\phi$ can be decomposed as
\be
\phi_j^{(i)} = \chi_j^{(i)}(x_W,x_H) + \tilde\chi_j^{(i)}(\{x_i,y_a\})
\ee
for the parts involving quarks and bosons and the parts with 
squarks and charginos, respectively. The functions $\chi_j^{(i)}(x_W,x_H)$
have been calculated already \cite{Bur90,Urb98}, here we just
restate the result
\bea
\chi^{(1)}_{LL}(x_W,x_H) &=& 
L^{(1,THDM)}(x_W,xH)  \nonumber\\ 
&& + \left[2\xi\left(1+g_{IR}+\ln(x_{\mu_0})\right)+6\ln(x_{\mu_0})\sum_{i=H,W}
           \frac{x_i\partial}{\partial x_i}\right]\,S(x_W,x_H)\,, \nonumber\\
\chi^{(8)}_{LL}(x_W,x_H) &=& 
L^{(8,THDM)}(x_W,xH)  \nonumber\\ 
&& + \left[2\xi\left(1+g_{IR}\right)+(3+\xi)\ln(x_bx_d)\right]\,S(x_W,x_H) 
\,,\nonumber\\
\chi_1^{(8)}(x_W,x_H) &=& -(3+\xi)\,S(x_W,x_H)
\,,\nonumber\\
\chi_2^{(8)}(x_W,x_H) &=& -2\chi_3^{(1)}(x_W,x_H) =
-(3+\xi)\frac{m_bm_d}{m_d^2-m_b^2}\ln
\left(\frac{x_b}{x_d}\right)\,S(x_W,x_H)\,.
\eea
All other $\chi$ equal zero. We have introduced here some other abbreviations,
\be
x_{\mu_0} = \frac{{\mu_0}^2}{M_W^2}\;,\;
g_{IR}=-\frac{x_d\ln(x_d)-x_b\ln(x_b)}{x_d-x_b}\,,
\ee
where ${\mu_0}$ is the scale of integrating out the heavy degrees of freedom.
Note, that all masses entering the functions are the masses taken at this 
scale, for example, $m_t$ is the top--mass renormalized within the
$\ol{MS}$--scheme at the scale ${\mu_0}$.

For the squarks and charginos running in the boxes the functions
$\tilde\chi$ differ only slightly from the $\chi$ just presented.
Namely, for the $\tilde\chi_{1,2,3}$ one just has to replace 
$S(x_W,x_H)$ by $\tilde S(\{x_i,y_a\})$. The $\tilde\chi_{LL}$
read for $\xi = 1$
\bea
\tilde\chi^{(1)}_{LL}(\{x_i,y_a\}) &=& 
\sum\limits_{i,j=1}^2\sum\limits_{a,b=1}^6 \tilde K_{ij,ab}
\left\{\tilde{L}^{(1)}(x_{i,j},y_{a,b})\vphantom{\frac{|}{|}}
+ \left[2\left(1+g_{IR}+\ln(x_{\mu_0})\right)
\right.\right.
\nonumber\\
&&\indthree +\left.\left.
        2\ln(x_{\mu_0})\sum_{k=i,j}\sum_{c=a,b}\left(
         \frac{3x_i\partial}{\partial x_i}+\frac{2y_c\partial}{\partial y_c}
         \right)\right]
\right\}\,S(x_{i,j},y_{a,b}) \,,\nonumber\\
\tilde\chi^{(8)}_{LL}(\{x_i,y_a\}) &=& 
\sum\limits_{i,j=1}^2\sum\limits_{a,b=1}^6 \tilde K_{ij,ab}
\left\{\tilde{L}^{(8)}(x_{i,j},y_{a,b})
\right.
\nonumber\\
&&\indthree \left.
+ \left[2\left(1+g_{IR}\right)+4\ln(x_bx_d)\right]
\right\}\,S(x_{i,j},y_{a,b})\,.
\eea
To perform the calculations we used Mathematica \cite{mathem}
and the Mathematica--based package FeynArts \cite{FArts}. 

Note that the results so far show a non--trivial dependence on the
gauge parameter and the IR--masses. This dependence will vanish after
we matched the full and effective sides of the theory. The functions 
$L$ constitute an integral part of the Wilson--coefficient emerging 
after this procedure, and are gauge and IR--independent.
They are listed in the Appendix.

\subsection{Matching and running}

To match the full and the effective side of the theory we have to evaluate
the matrix element of the physical operator $\OLL$ up to order $O(\alpha_s)$
using the same regularization, renormalization and gauge prescriptions 
employed before. Yet, the one--loop amplitude of $\OLL$ as given by
Fig.\,\ref{Figeta2} results in the same operators with the same coefficients 
for the obviously unphysical operators ${\cal O}_{1,2,3}$,
\bea\label{OLLmu}
\langle \OLL({\mu_0})\rangle_{1\rm loop} &=&
\langle \OLL({\mu_0})\rangle_{\rm tree} \nonumber\\
&&+ \frac{\alpha_s({\mu_0})}{4\pi}\,
\sum_k\left(C_F\chi_{\Delta,k}^{(1)}({\mu_0})+\tilde{C}_A\chi_{\Delta,k}^{(8)}({\mu_0})\right)
{\bf 1}\otimes{\bf 1}\,\langle{\cal O}_k\rangle_{\rm tree}\,,
\eea
where again the sum over $k=LL,1,2,3$ is understood. 
We have used the identity
\be
T_a\otimes T^a = \tilde{C}_A{\bf 1}\otimes{\bf 1}
\ee
to allow for rearranging the colours under Fierz--transformations.
Now the $\chi_{\Delta}({\mu_0})$ read
\bea
\chi^{(1)}_{\Delta,LL}({\mu_0}) &=&  -3 + 2\xi(1+\ln(x_{\mu_0})+g_{IR})
\,,\nonumber\\
\chi^{(8)}_{\Delta,LL}({\mu_0}) &=&  -5 + 2\xi(1+g_{IR})+(3+\xi)\ln(x_bx_d)
                              -6\ln(x_{\mu_0})
\,,\nonumber\\
\chi^{(8)}_{\Delta,1}({\mu_0}) &=& -(3+\xi)
\,,\nonumber\\
\chi^{(1)}_{\Delta,3}({\mu_0}) &=& \frac12 \chi^{(8)}_{\Delta,2}({\mu_0})
=\frac{3+\xi}{2}\frac{m_bm_d}{m_d^2-m_b^2}\ln\left(\frac{x_b}{x_d}\right)\,,
\eea
and equation (\ref{OLLmu}) enables us to write the matrix element of the 
effective Hamiltonian up to order $O(\alpha_s)$ as
\be\label{heff1}
\langle H_{\rm eff}\rangle = 
\langle H_{\rm eff}^{(0)} + \Delta H_{\rm eff}^{(1)}\rangle = 
\frac{G_F^2}{4\pi^2}M_W^2(V_{td}V_{tb}^\ast)^2\,C_{LL}({\mu_0})
\langle \OLL({\mu_0})\rangle_{\rm 1loop}\,.
\ee
We are now able to expand the Wilson--coefficient $C_{LL}$ up to 
$O(\alpha_s)$,
\be\label{Wilson1}
C_{LL}({\mu_0}) = S + \frac{\alpha_s}{4\pi}\,D
\ee
with $S$ given by equation (\ref{S1}). Comparing coefficients yields the
result for $D$,
\bea
D &=& D(x_W,x_H,x_{\mu_0}) + \tilde D(\{x_i,y_a\},x_{\mu_0})
\,,\nonumber\\
D(x_W,x_H,x_{\mu_0}) &=& 
\;\;\;C_F\left\{L^{(1,THDM)}(x_W,xH) + 
      \left[3+6\ln(x_{\mu_0})\sum_{i=H,W}
           \frac{x_i\partial}{\partial x_i}\right]\,S(x_W,x_H)\right\}
 \nonumber\\
&& + \tilde{C}_A\left\{L^{(8,THDM)}(x_W,xH) +
     \left[5+6\ln(x_{\mu_0})\right]\,S(x_W,x_H)\right\}
\,,\nonumber\\
\tilde D(\{x_i,y_a\},x_{\mu_0}) &=& 
\;\;\; C_F\sum\limits_{i,j=1}^2\sum\limits_{a,b=1}^6 \tilde K_{ij,ab}
\left\{\tilde{L}^{(1)}(x_{i,j},y_{a,b})\vphantom{\frac{|}{|}}
\right.
\nonumber\\
&&\indone \left.
+ \left[3+6\ln(x_{\mu_0})\sum_{k=i,j}\sum_{c=a,b}\left(
         \frac{x_i\partial}{\partial x_i}+\frac{y_c\partial}{2\partial y_c}
         \right)
\right]\,S(x_{i,j},y_{a,b})\right\}
 \nonumber\\
&&
+ \tilde{C}_A\sum\limits_{i,j=1}^2\sum\limits_{a,b=1}^6 \tilde K_{ij,ab}
\left\{\tilde{L}^{(1)}(x_{i,j},y_{a,b})\vphantom{\frac{|}{|}}
+ \left[5+6\ln(x_{\mu_0})\right]\,S(x_{i,j},y_{a,b})\right\}
\,.\nonumber\\
\eea
Hence, at this stage we have expressions for the Wilson--coefficients at the
matching scale $\mu_0$, the last step to be performed is the evolution down 
to the mesonic scales.

This evolution is achieved with help of the renormalization group.
The solution of the renormalization group equation for the Wilson--coefficient
$C_{LL}$ at NLO reads
\be
C_{LL}(\mu) = \eta_{2,LO}\left[1+
\frac{\alpha_s(M_W)-\alpha_s(\mu)}{4\pi}\,Z_5\right]\,C_{LL}(M_W)
\ee
with the matching scale chosen explicitly as $\mu_0 = M_W$. The term 
$Z_{n_f}$ is given along with other ingredients needed as
\bea
\beta^{(1)}_{n_f} &=& 
\frac{34}{3}\,N_c^2 \,-\, \frac{10}{3}\,N_cn_f \,-\, 2C_fn_f
\,,\nonumber\\
\gamma^{(1)}_{n_f} &=&
\frac{N_c-1}{2N_c}\left[-21\,+\,\frac{57}{N_c}\,-\,\frac{19}{3}\,N_c
\,+\,\frac43\,n_f\right]
\,,\nonumber\\
Z_{n_f} &=& \frac{\gamma^{(1)}_{n_f}}{2\beta^{(0)}_{n_f}}
-\frac{\gamma^{(0)}}{2{\beta^{(0)}_{n_f}}^2}\beta^{(1)}_{n_f}\,.
\eea
The leading order expressions have been given in Eqs.\,(\ref{etaLO},
\ref{bgLO}).
Defining
\be\label{opredef}
\tilde{\cal O}_{LL} = \alpha_s(\mu)^{-g^{(0)}/(2\beta^{(0)}_{n_f})}
\left[1+\frac{\alpha_s(\mu)}{4\pi}\,Z_{n_f}\right]\,\OLL(\mu)\,.
\ee
and setting $n_f=5$ for the $B$--system
we can cast the effective Hamiltonian for $\Delta B = 2$--transitions at the 
mesonic scale in the following form
\be
H_{\rm eff} = \frac{G_F^2}{4\pi^2}M_W^2(V_{td}V_{tb}^\ast)^2
\eta_2 \,S\, \tilde{\cal O}_{LL}
\ee
where the factor $\eta_2$ describing QCD--corrections up to 
Next--to--Leading Order is determined by
\be
\eta_2 = \alpha_s(M_W)^{g^{(0)}/(2\beta^{(0)}_{n_f})}
\left[1+\frac{\alpha_s(M_W)}{4\pi}\left(\frac{D}{S}+Z_{n_f}\right)\right]
\ee
with $n_f = 5$. Obviously neither the factor $\eta_2$ nor the redefined
operator $\tilde{\cal O}_{LL}$ depend on the energy scale $\mu$,
the formal dependence within the definition of the operator is compensated by
the Bag--parameter as given in Eq.\,(\ref{Bagdef}).

\subsection{Generalization on $\eta_{2K}$} 

We want to generalize the considerations leading to an expression of
$\eta_2$ for the $B$--system now on the $K$--system. Actually this 
reduces to the question of how to incorporate the effect of the
thresholds on the renormalization group evolution. 
In LO this can be seen easily in Eq.\,(\ref{etaLO}), and a generalization
is straightforward. For $\eta_{2K}$ we obtain the following result up to
NLO
\bea
\eta_{2K} &=& \alpha_s(m_c)^{g^{(0)}/(2\beta^{(0)}_{3})}
\left(\frac{\alpha_s(m_b)}{\alpha_s(m_c)}\right)^{g^{(0)}/(2\beta^{(0)}_{4})}
\left(\frac{\alpha_s(M_W)}{\alpha_s(m_b)}\right)^{g^{(0)}/(2\beta^{(0)}_{5})}
\nonumber\\
&&\left[1+\frac{\alpha_s(m_c)}{4\pi}(Z_3-Z_4)
         +\frac{\alpha_s(m_b)}{4\pi}(Z_4-Z_5)
+\frac{\alpha_s(M_W)}{4\pi}\left(\frac{D}{S}+Z_{5}\right)\right]\,.
\eea
So we are now able to derive an expression for the effective Hamiltonian
for $\Delta S = 2$--transitions including QCD corrections up to
$O(\alpha_s)$.
\bea
\label{heffK1}
H_{\rm{eff}} &=& \frac14
    \,{G^2_F \over \pi^2}\; m_W^2 \;
    \left[{( V_{cd}V^\ast_{cs})}^2 \eta_1\,S(x_c) + 
          {( V_{td}V^\ast_{ts})}^2 \eta_2\,S  \right.
\nonumber\\&&\indtwo\indone \left.
     +(V_{cd}V^\ast_{cs})( V_{td}V^\ast_{ts}) \eta_3\,S(x_c,x_W) 
     \right]\;\tilde{\cal O}_{LL}\,.
\eea
Here, the operator $\tilde{\cal O}_{LL}$ is defined by equation
(\ref{opredef}) with $n_f=3$.

\subsection{Gluino mediated corrections}

As should be noted, there are corrections of the various box--diagrams
mediated by gluinos as well. Basically these corrections exhibit the same
topological structure as the gluon--corrections, but to some extend they
mix the sectors constituting the Flavour Changing Neutral Currents,
resulting for example in boxes containing $W$--bosons, charginos, quarks 
and squarks (this is achieved by replacing the gluon of topology
g of Fig.\,\ref{FigTHDMNLO}). In addition to the non--vanishing mass
off the gluino this definitely complicates the computational
situation considerably.

Nevertheless, \cite{Ciu98} clearly shows, that the vertex--corrections
due to the gluinos are of great importance and do not vanish in the limit
of a heavy gluino--mass. In other words, the gluino does not decouple 
from an effective theory containing only the lighter degrees of
freedom. In our opinion this indicates that more care has to be taken
in dealing with the gluino corrections in a consistent fashion.

\section{Results}

We want to compare now some very well measured experimental quantities
related to the effective Hamiltonians we investigated.

$B\ol B$--mixing can be described by the mass difference $\Delta M$ of the two
mass eigenstates, given by
\bea
\Delta M &=& 
\frac{|\langle \bar{B^0}|\,H_{\rm eff}(\Delta B = 2)\,|B\rangle|}{m_B}
\nonumber\\
&=& 
\frac{G_F^2}{6\pi^2} M_W^2\,(V_{td}V_{tb}^\ast)^2\,\eta_2\,S\,f_B^2 m_B B_B\,,
\eea
where we have employed equation (\ref{Bagdef}).

Similar expressions hold for the $K$--system. For example, the parameter 
$\epsilon_K$ for indirect CP--violation in the decay $K\to\pi\pi$ is very 
well approximated by 
\bea
\epsilon_K &=& \frac{G_F^2}{6\pi^2} \frac{F_K^2 B_K m_K}{\sqrt{2}\Delta M_K}
               M_W^2 Im[V_{td}V_{ts}^*]
\nonumber\\
&&\left\{Re[V_{cd}V_{cs}^*]\,[\eta_1 S(x_c) - \eta_S(x_c,x_W)] -
Re[V_{td}V_{ts}^*]\,\eta_{2K}\,S\right\}\,,
\eea
under the use of the relation $Im[V_{cd}V_{cs}^*]^*=Im[V_{td}V_{ts}^*]$
\cite{Buc96}.

\subsection{Inputs}

Before we present our results, we want to list our input parameters.
Note, that we use the Wolfenstein--parametrization with parameters
$A, \lambda, \rho, \eta$. We start with the SM--parameters and
experimental data. They can be found in \cite{PDG}.
\bea
\alpha_s(M_W) &=& 0.12\;,\; \alpha_s(m_b) = 0.22 \;,\; \alpha_s(m_c) =  0.34
\,,\nonumber\\
A &=& 0.80 \pm 0.08 \;,\; \lambda = 0.22
\,,\nonumber\\
m_t(M_W) &=& 167 \pm 10 \,\mbox{\rm GeV}\;,\;
\nonumber\\
\sqrt{B_B f_B^2} &=& 0.2\pm 0.04 \,\mbox{\rm GeV}\;,\;
\sqrt{B_K f_K^2} = 0.135\pm 0.015 \,\mbox{\rm GeV}
\,,\nonumber\\
\epsilon_K &=& 2.26\pm0.02 \cdot 10^{-3}\;,\;
\nonumber\\
\Delta M_B &=& 3.1\pm 0.1 \cdot 10^{-13} \,\mbox{\rm GeV}\;,\;
\Delta M_K = 3.5 \cdot 10^{-15} \,\mbox{\rm GeV}\;.
\eea
For the MSSM we made, as stated before, some assumptions concerning the
parameters. First of all, we want to neglect all flavour changing entries 
in the squark mixing matrix. This leaves us with the following form of the
two matrices $Z$ entering the final expressions
\be\label{MSSMmatrices}
Z_U = \left(
\begin{array}{rrrrrr}
1&0&0&0&0&0\\
0&1&0&0&0&0\\
0&0&\cos\phi&0&0&\sin\phi\\
0&0&0&1&0&0\\
0&0&0&0&1&0\\
0&0&-\sin\phi&0&0&\cos\phi\\
\end{array}\right)\;,\;
Z^+ = \left(
\begin{array}{rr}
\cos\chi&\sin\chi\\
-\sin\chi&\cos\chi\\
\end{array}\right)\;.
\ee
We dealt with the mixing angles $\phi$ and $\chi$ as free parameters
in the ranges $[-\pi/4,\pi/4]$.
We further considered for the masses of the squarks and the charginos 
the ranges
\bea
m_{\tilde u_{\{1,2,4,5\}}} &=& m_{\tilde q}\;,\;
m_{\tilde u_3} = m_{\tilde t_L}\;,\;
m_{\tilde u_6} = m_{\tilde t_R}\;,
\nonumber\\
150\,\mbox{\rm GeV} &\leq& m_{\tilde t_R} \leq 300\,\mbox{\rm GeV}\;,\;
m_{\tilde t_R} \leq m_{\tilde t_L} \leq 600\,\mbox{\rm GeV}\;,\;
m_{\tilde t_L} \leq m_{\tilde q} \leq 900\,\mbox{\rm GeV}\;,
\nonumber\\
100\,\mbox{\rm GeV} &\leq& m_{\tilde \chi_2},\ m_{\tilde \chi_1} 
\leq 400\,\mbox{\rm GeV}\;.
\eea
All references on a ``reduced'' SUSY denote another assumption, namely
only one active squark and chargino. In other words, in this case
we take $\phi=\chi=0$ and $\chi_1$ heavy enough to give no more
sizeable effects, e.g. decoupled. If not stated otherwise we took the
central values of the SM--parameters. 

\subsection{Results for the $B$--system and $\epsilon_K$}

It should be noted here, that for the $B_s$--system
there are so far only lower bounds on the size of the mass--splitting.
They are still somewhat lower than what can be expected from a theoretical point
of view within the SM. Because the models we investigated here
interfere additively with the SM we did not explicitly show any figures for
the $B_s$--system. Of course our results for $\eta_2$ and the mass--splitting
hold as well, as long as one replaces the flavour $d$ with $s$ when
needed.

\begin{figure}\begin{center}
\,\mbox{\epsfxsize=12.0cm\epsfysize=8.0cm\epsffile{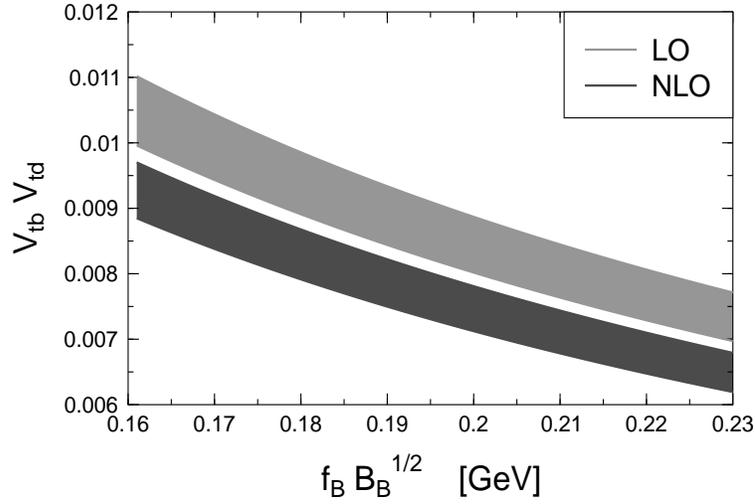}}
\parbox[t]{8cm}{\caption{\label{ckmbb}\footnotesize 
The product of the CKM--elements 
$|V_{td}V_{tb}^\ast|$ versus $\sqrt{B_B f_B^2}$ for the allowed region of 
$m_t$ within the SM.}}
\end{center}\end{figure}

\begin{figure}\begin{center}
\mbox{\epsfxsize=12.0cm\epsfysize=8.0cm\epsffile{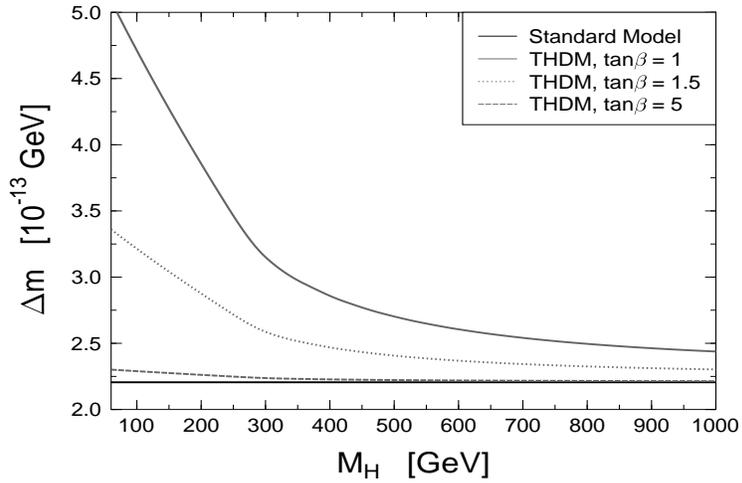}}
\parbox[t]{8cm}{\caption{\label{mhtan}\footnotesize 
The influence of a pure THDM for
different values of $M_H$ and $\tan\!\beta$. We chose $\sqrt{B_B f_B^2} = 
0.18\,\mbox{\rm GeV}$, $|V_{td}V_{tb}^\ast| = 0.007$ and 
$m_t(M_W) = 167 \,\mbox{\rm GeV}$.}}
\end{center}\end{figure}

\begin{figure}\begin{center}
\mbox{\epsfxsize=12.0cm\epsfysize=8.0cm\epsffile{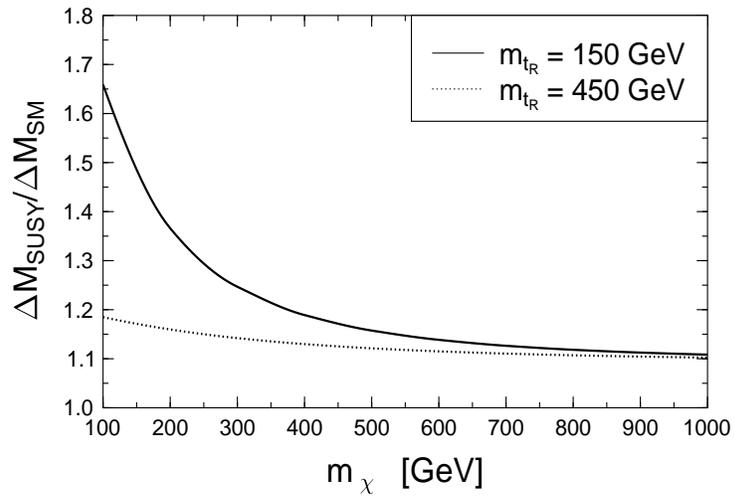}}
\parbox[t]{8cm}{\caption{\label{susyrel}\footnotesize 
The ratio of the mass--splittings 
$\Delta M$ obtained for our ``reduced'' MSSM and the SM. We assumed a 
Higgs-mass of $M_H = 500\,\mbox{\rm GeV}$ and $\tan\!\beta = 1.5$.}}
\end{center}\end{figure}

Obviously, as can be deduced from Fig.\,\ref{ckmbb}, the large range for
the hadronic parameters of the $B$--system allows for a large range
of the CKM--elements $|V_{td}V_{tb}|$ responsible for the mixing even in 
the SM. Furthermore, one can read of the effect of NLO--corrections
in comparison to the LO result to be of the order of roughly $10\%$.
Fig.\,\ref{mhtan} clearly shows the possibly large influence of the 
charged Higgs--sector on the mixing within the $B$--system. Obviously
this sets some constraints on Higgs--parameters within the THDM or even the 
MSSM, because the supersymmetric particles interfere additively as well.
This is displayed in Fig.\,\ref{susyrel}, where we have used our 
``reduced'' MSSM--parameter space. This choice provides a good feeling
for the effect of the MSSM, by the way, and we can read of possibly large 
effects on the mixing induced by the MSSM. Stated the other way around,
the MSSM clearly has the potential to account for sizeable effects on the
CKM--elements, pushing them to lower values. Note, that similar 
statements hold for the $K$--system as well, because the contribution of the 
$c$--quarks is not the dominant one.

In the other plots, the scatterplots, we have displayed the effect of 
NLO--corrections to $\eta_2$ as compared to the LO--result and found
an effect of roughly $10\%$, the same as in the SM.

Furthermore, we compared the effect of taking into account the full MSSM using 
the GIM--mechanism for the first two generations within the ranges given 
above. Clearly, the MSSM has the effect of altering the result obtained
for the SM. However, as the plots show, the effect of a not decoupled
charged Higgs is sizeable, and in this sense models like the THDM 
are worth considering.

\subsection{Effects on the unitarity triangle}

\begin{figure}\begin{center}
\mbox{\epsfxsize=12.0cm\epsffile{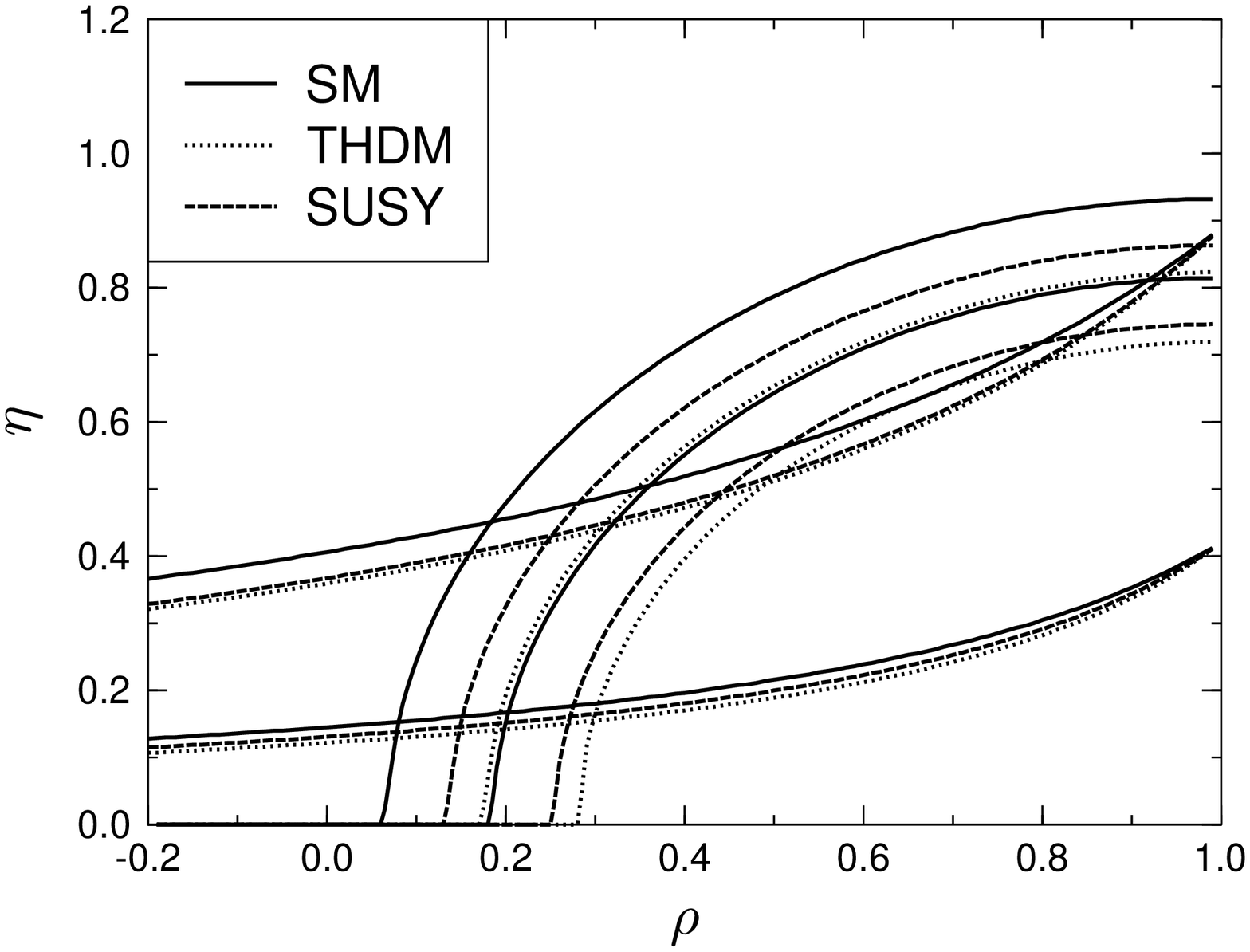}}
\parbox[t]{8cm}{\caption{\label{rhoeta}\footnotesize 
The influence of adding THDM and MSSM to the SM
on the position of $(\rho,\eta)$. We varied the top--mass within the bounds 
indicated above as well as the hadronic parameters and chose 
$M_H=200\,\mbox{\rm GeV}$, $\tan\!\beta = 1.5$ for the THDM investigations 
and $M_H=500\,\mbox{\rm GeV}$, $\tan\!\beta = 5$,
$m_{\tilde t_R}=200\,\mbox{\rm GeV}$ and 
$m_{\tilde\chi_2}=120\,\mbox{\rm GeV}$
for the reduced MSSM.}}
\end{center}\end{figure}

As a matter of fact, the above considerations do influence the
investigations concerning the unitarity triangle. As shown by some
examples (see Fig.\,\ref{rhoeta}) it will be shifted by adding a THDM 
or a MSSM.

\section{Conclusions}

The investigations above yield a clear picture of the influence
of the MSSM on the mixing phenomena within the $B$-- and the
$K$--system. 

\begin{figure}\begin{center}
\begin{tabular}{c}
\mbox{\epsfxsize=12.0cm\epsfysize=6.0cm\epsffile{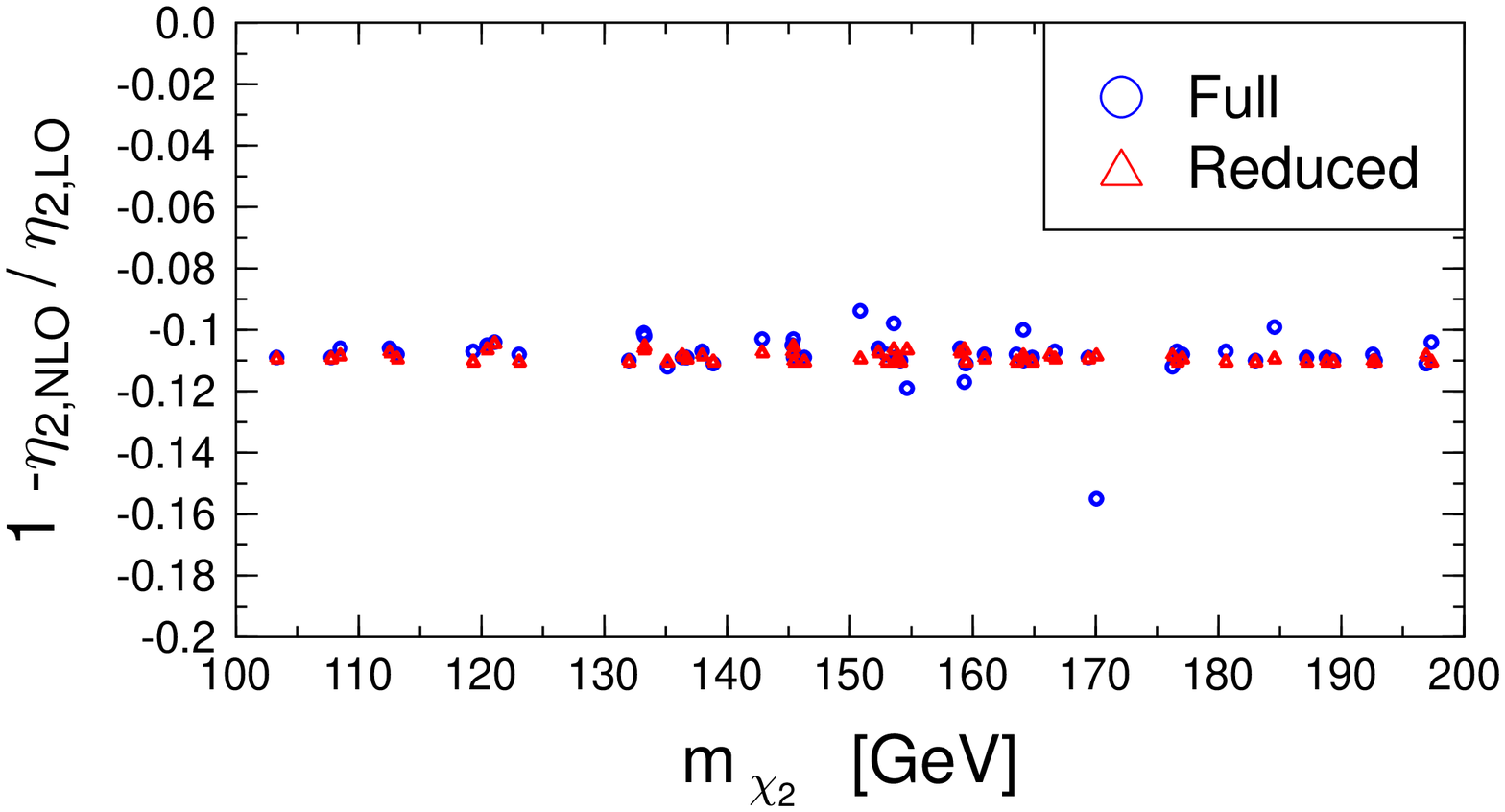}}\\
\mbox{\epsfxsize=12.0cm\epsfysize=6.0cm\epsffile{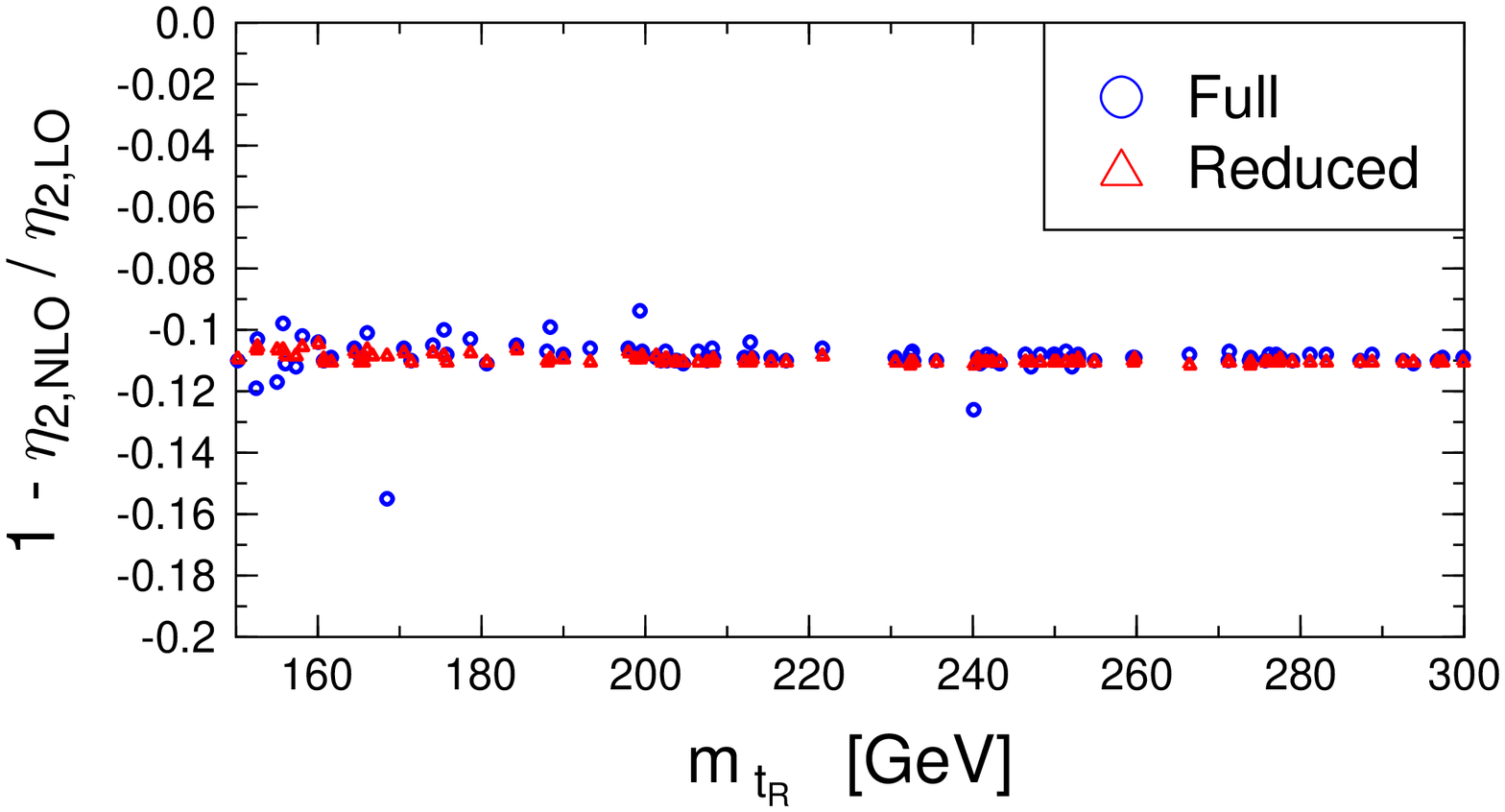}}\\
\mbox{\epsfxsize=12.0cm\epsfysize=6.0cm\epsffile{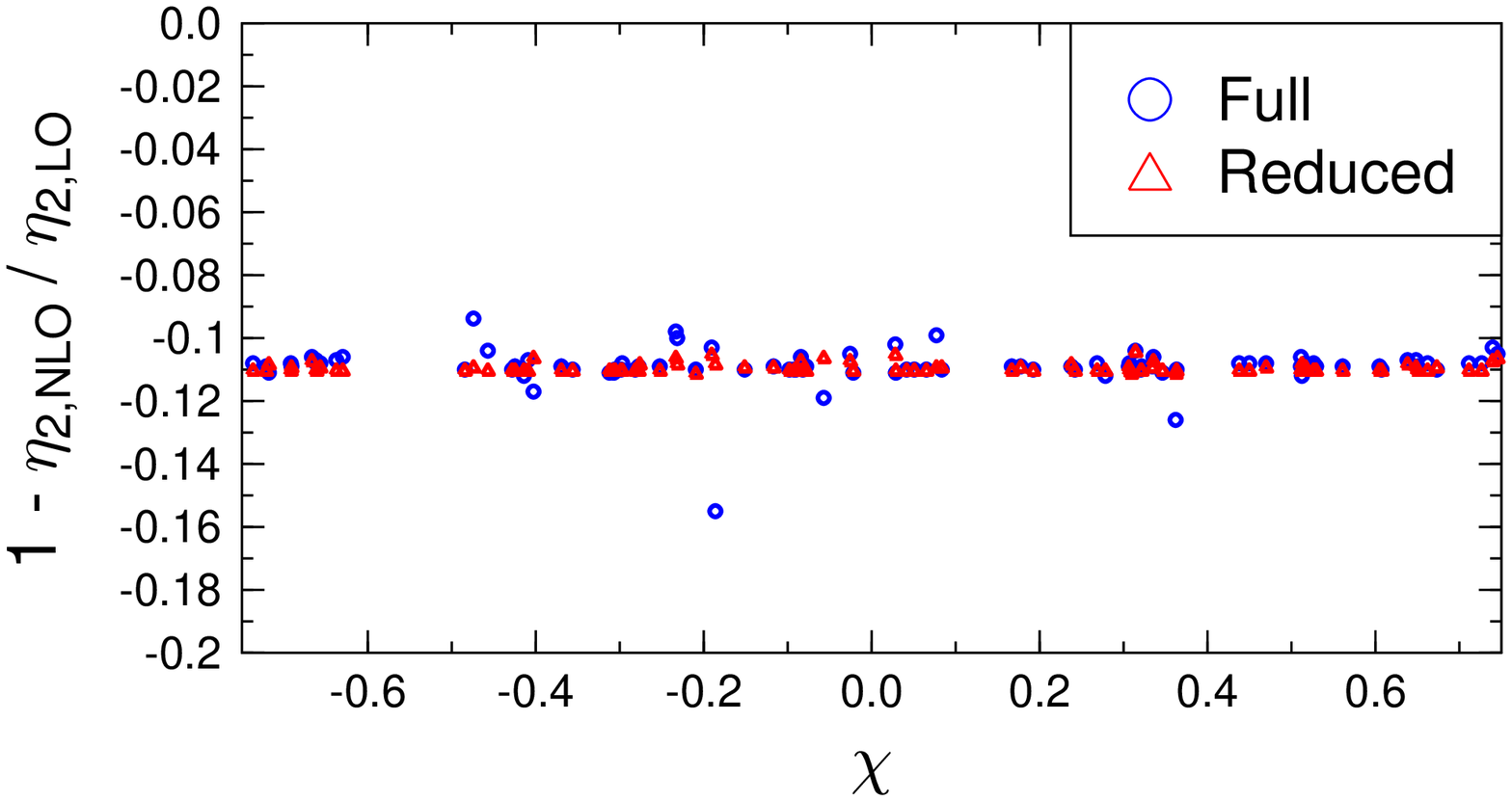}}
\end{tabular}
\parbox[t]{8cm}{\caption{\label{eta2}\footnotesize 
The ratio $\eta_{2LO}/\eta_{2NLO}$ in
dependence on $m_{\tilde t_R}$, $m_{\tilde\chi_2}$, $\chi$. In all cases
we practically decoupled the Higgs boson by setting 
$M_H = 1000\,\mbox{\rm GeV}$ and $\tan\!\beta = 5$. }}
\end{center}\end{figure}

Actually it doesn't seem as if NLO--corrections for the MSSM differ
considerably from the result obtained for the SM. The factor $\eta_2$ 
displayed in (\ref{eta2}) for our choice of the parameter space
seems to favour an effect of roughly $-10 \%$. However, the NLO
corrections do of course reduce the uncertainty involved with possibly large 
scale--differences, but it should be stressed that in case
these are too large, one has to perform a step--by--step procedure
when integrating out the heavy degrees of freedom. 

\begin{figure}\begin{center}
\begin{tabular}{c}
\mbox{\epsfxsize=12.0cm\epsfysize=6.0cm\epsffile{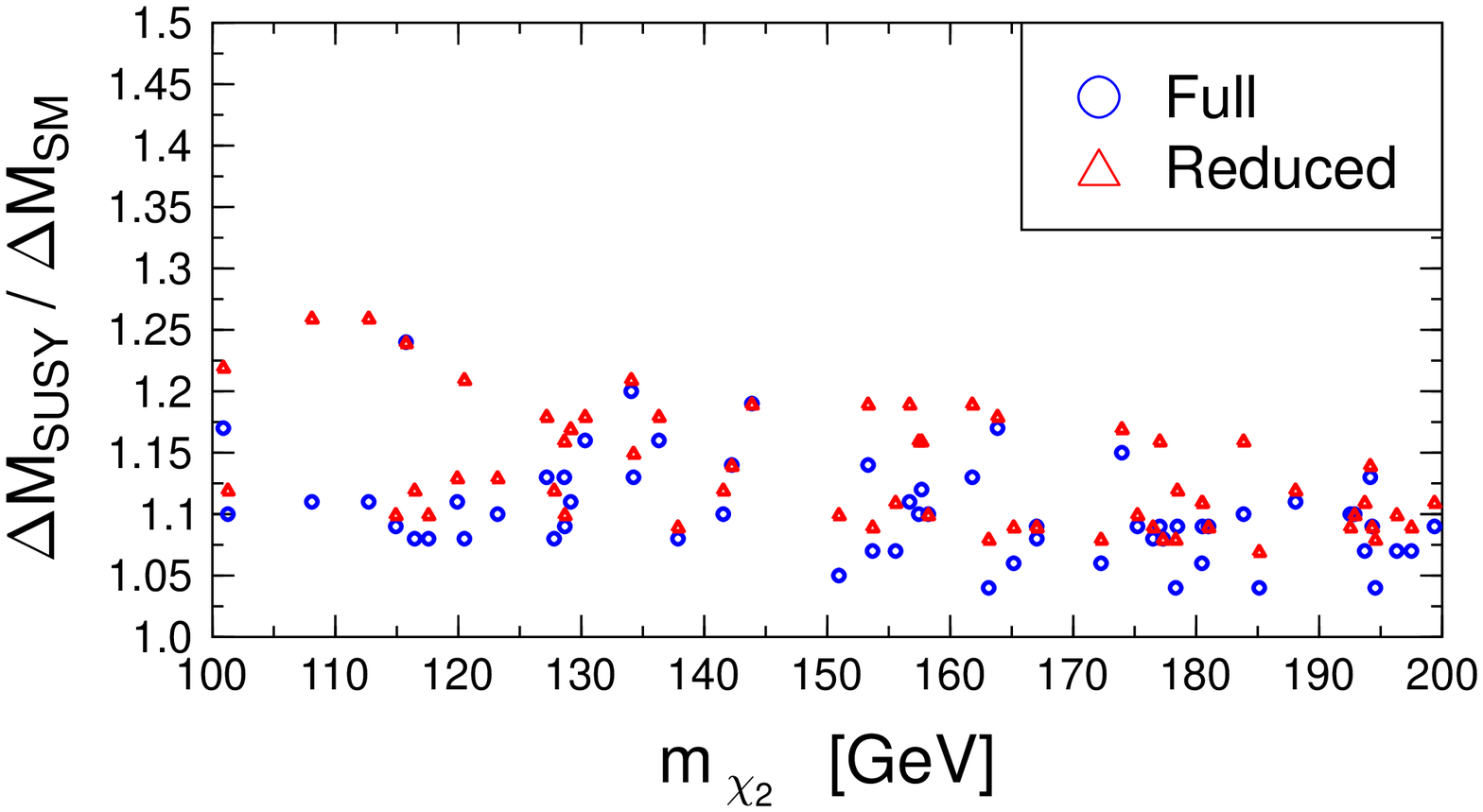}}\\
\mbox{\epsfxsize=12.0cm\epsfysize=6.0cm\epsffile{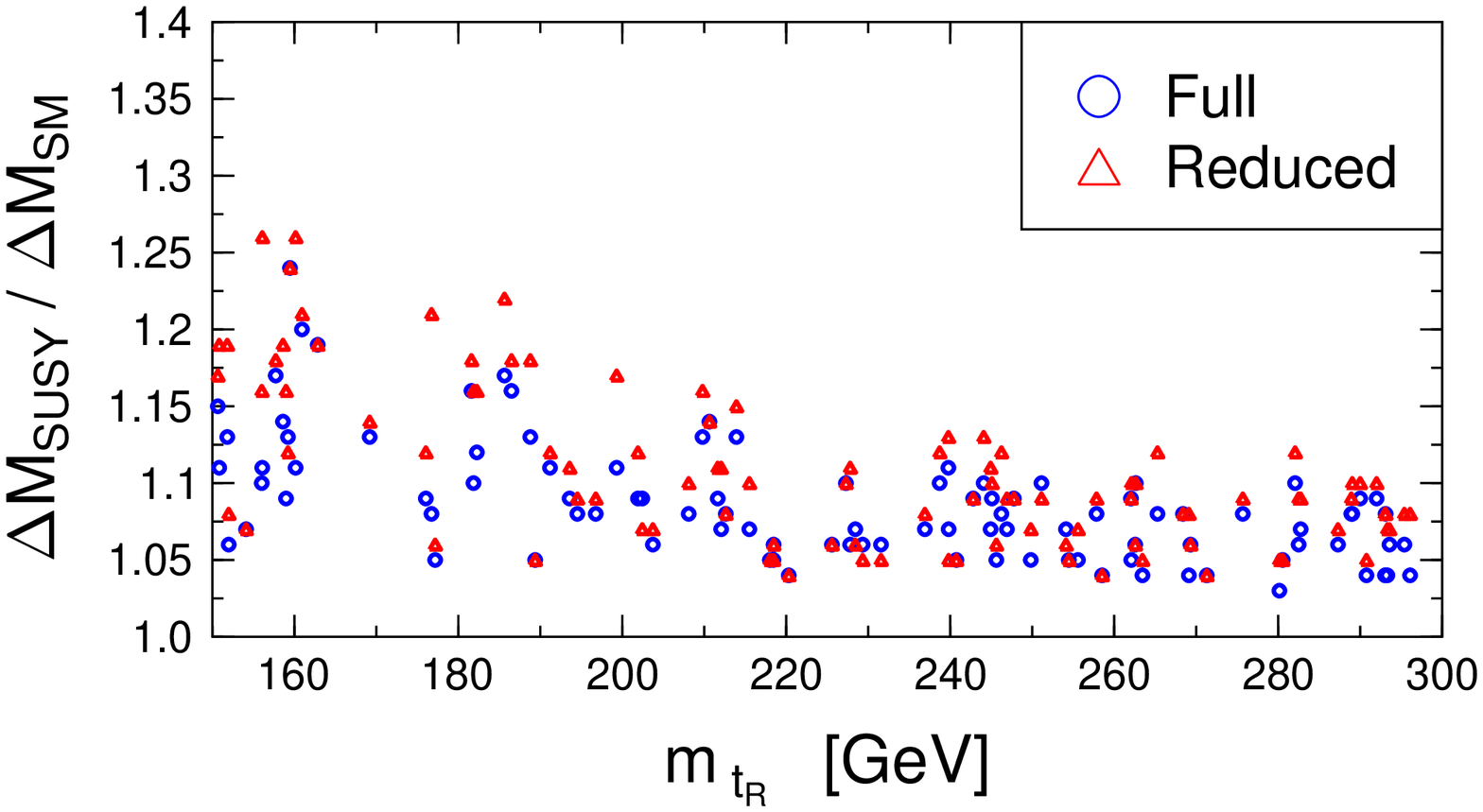}}\\
\mbox{\epsfxsize=12.0cm\epsfysize=6.0cm\epsffile{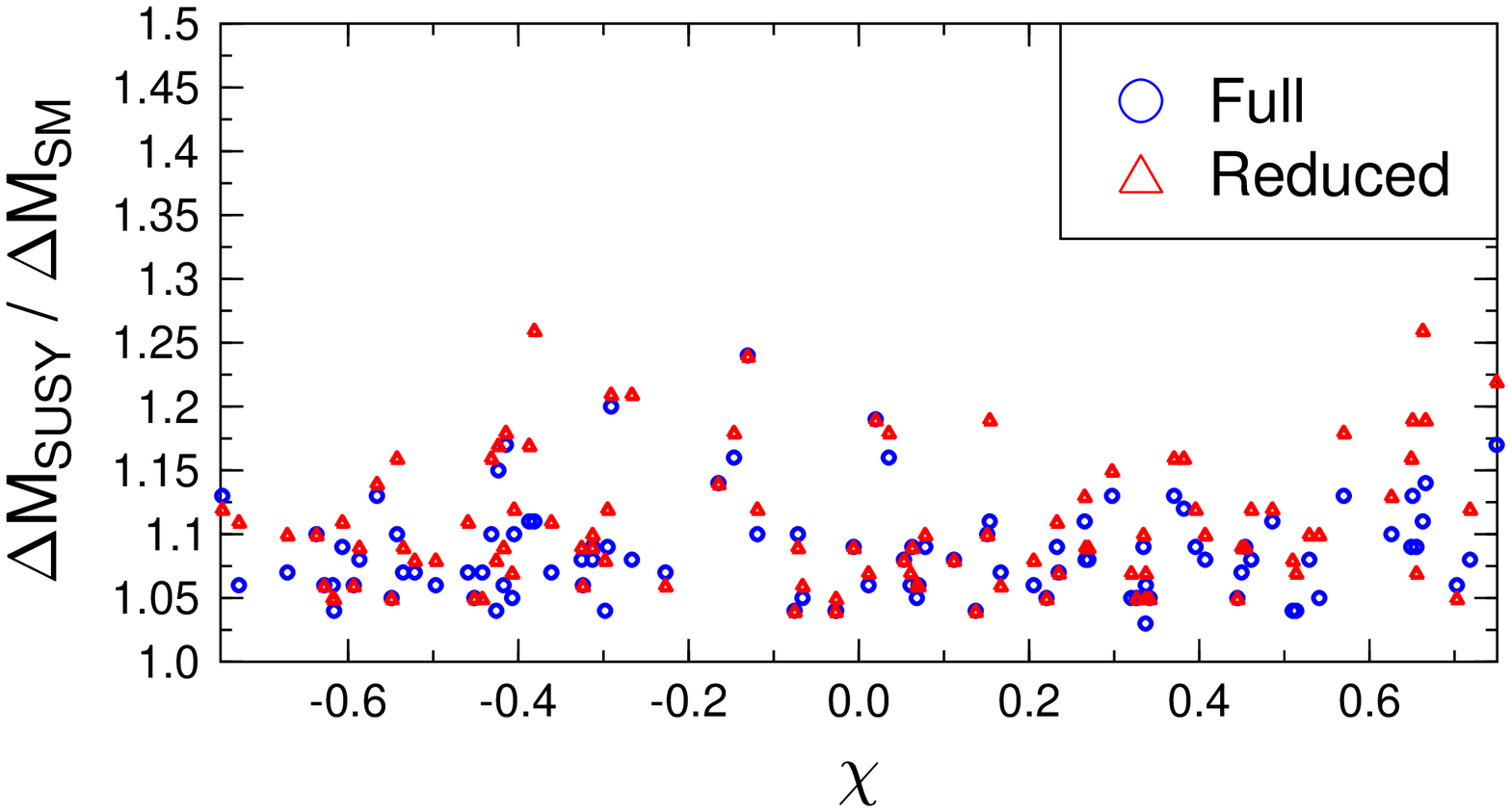}}
\end{tabular}
\parbox[t]{8cm}{\caption{\label{scatter1}\footnotesize 
The ratio of the 
mass--splittings $\Delta M$ within the MSSM and the SM in
dependence on $m_{\tilde t_R}$, $m_{\tilde\chi_2}$, $\chi$. In all cases
we practically decoupled the Higgs boson by setting 
$M_H = 1000\,\mbox{\rm GeV}$ and $\tan\!\beta = 5$. }} 
\end{center}\end{figure}

\begin{figure}\begin{center}
\begin{tabular}{c}
\mbox{\epsfxsize=12.0cm\epsfysize=6.0cm\epsffile{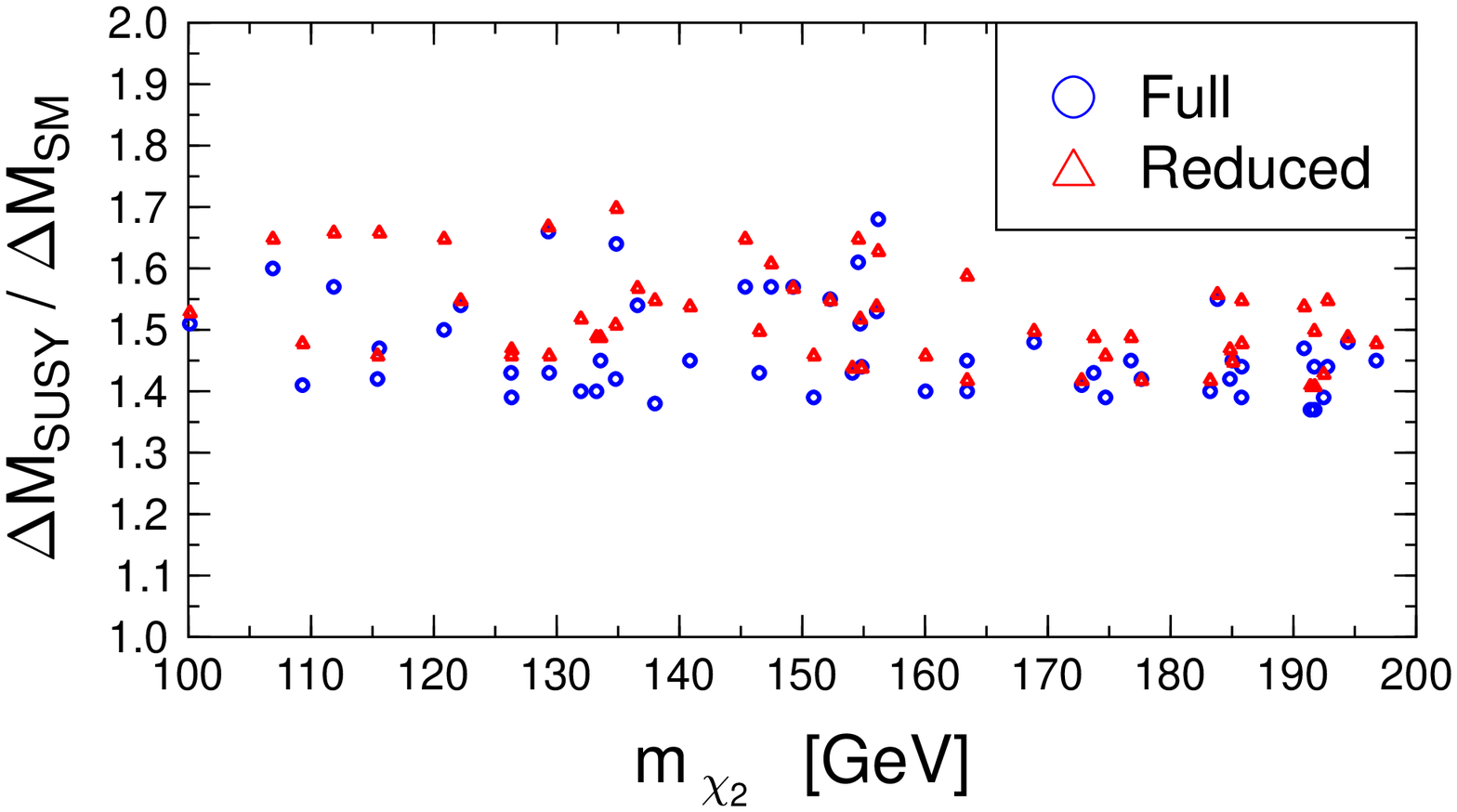}}\\
\mbox{\epsfxsize=12.0cm\epsfysize=6.0cm\epsffile{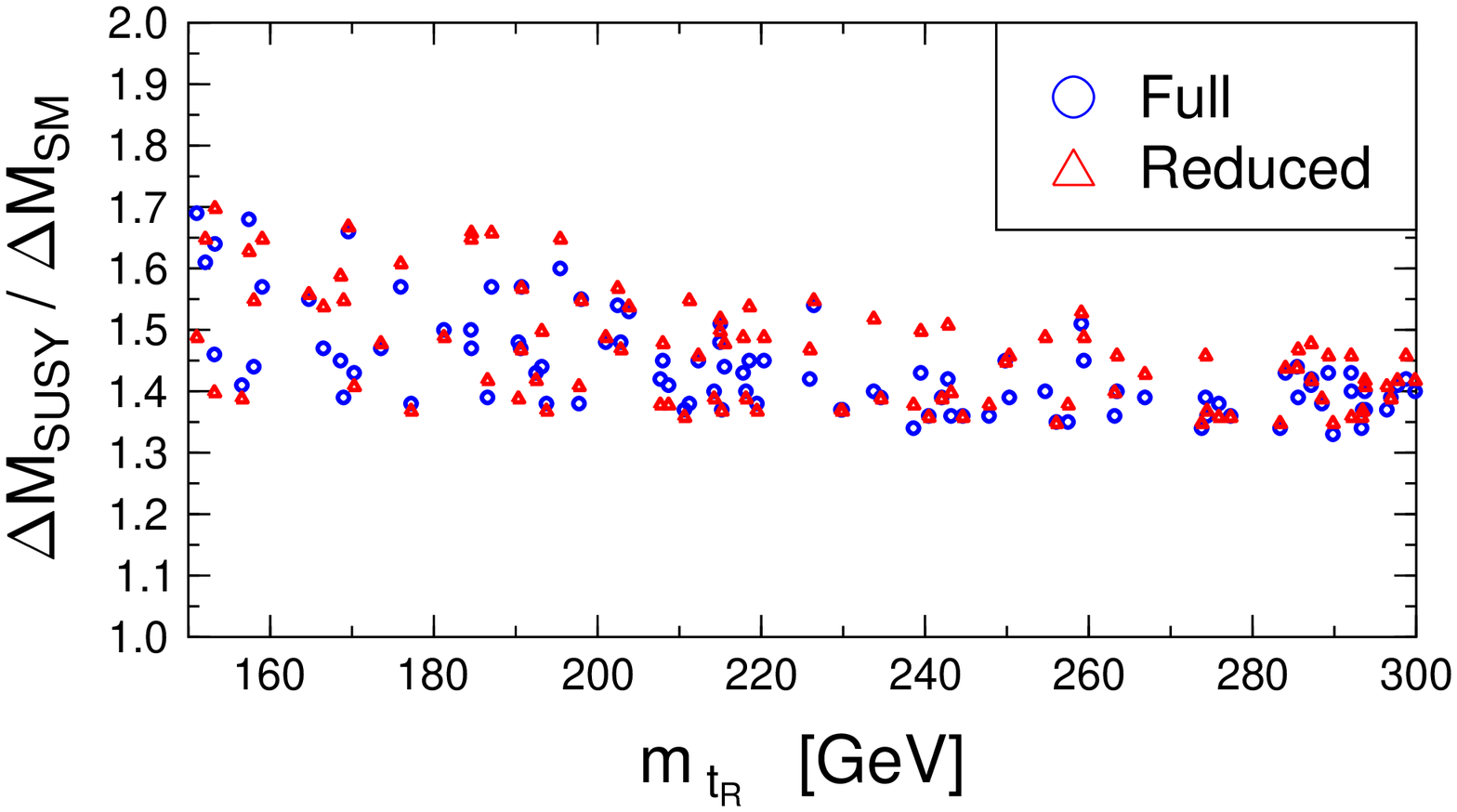}}\\
\mbox{\epsfxsize=12.0cm\epsfysize=6.0cm\epsffile{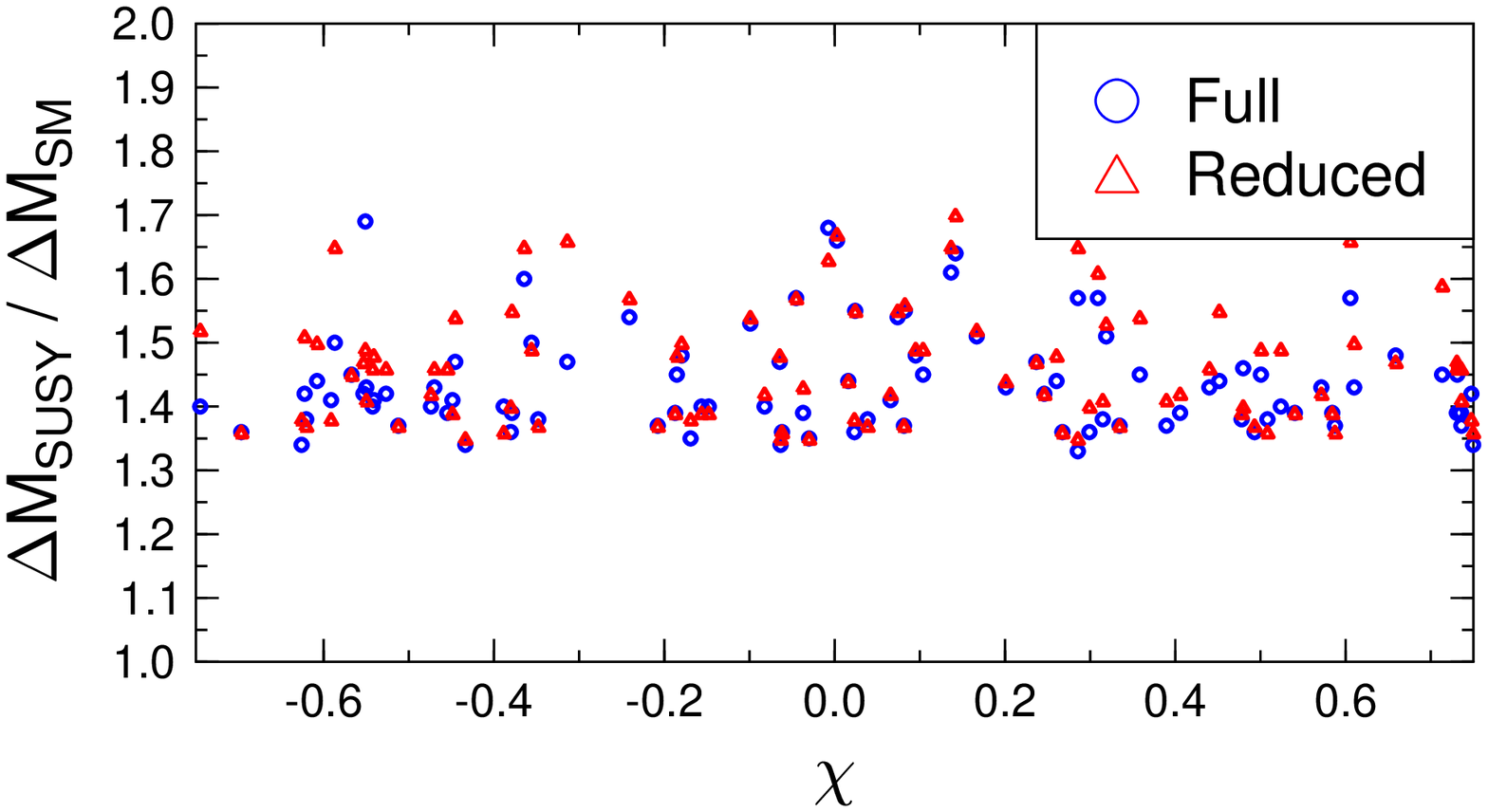}}
\end{tabular}
\parbox[t]{8cm}{\caption{\label{scatter2}\footnotesize 
The ratio of the 
mass--splittings $\Delta M$ within the MSSM and the SM in
dependence on $m_{\tilde t_R}$, $m_{\tilde\chi_2}$, $\chi$. In all cases
we included the Higgs boson by setting 
$M_H = 100\,\mbox{\rm GeV}$ and $\tan\!\beta = 1.5$. }} 
\end{center}\end{figure}

\begin{figure}\begin{center}
\begin{tabular}{c}
\mbox{\epsfxsize=12.0cm\epsfysize=6.0cm\epsffile{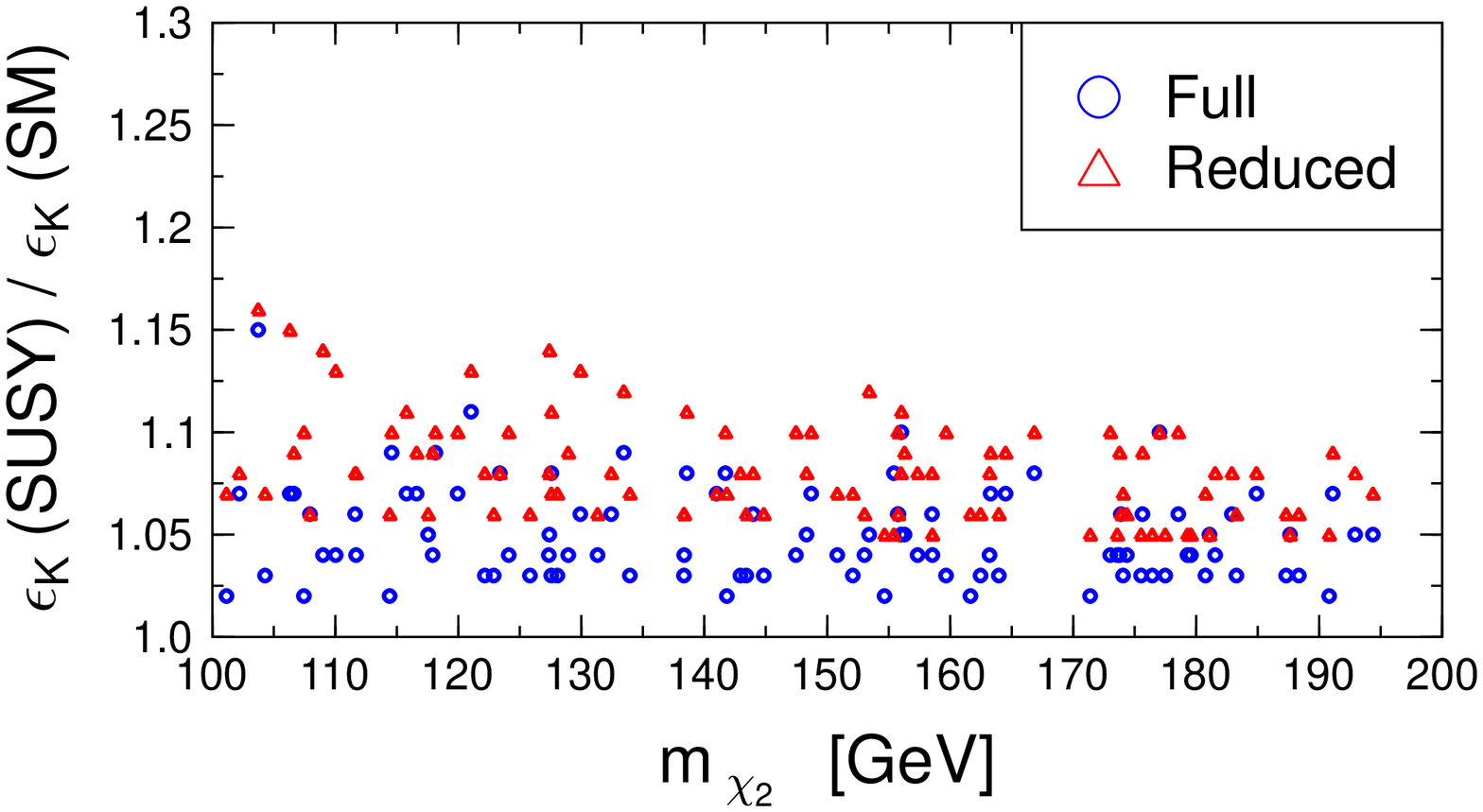}}\\
\mbox{\epsfxsize=12.0cm\epsfysize=6.0cm\epsffile{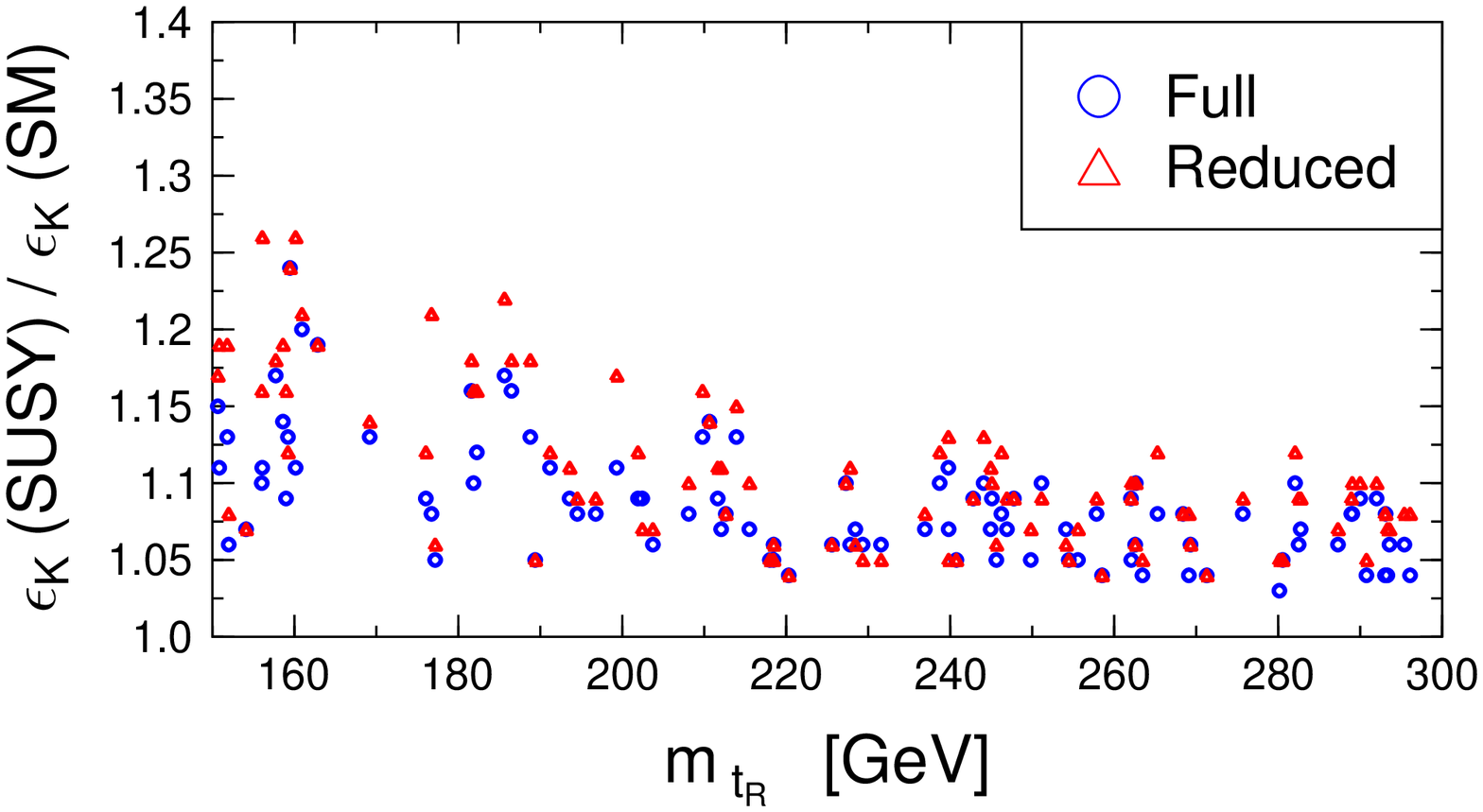}}\\
\mbox{\epsfxsize=12.0cm\epsfysize=6.0cm\epsffile{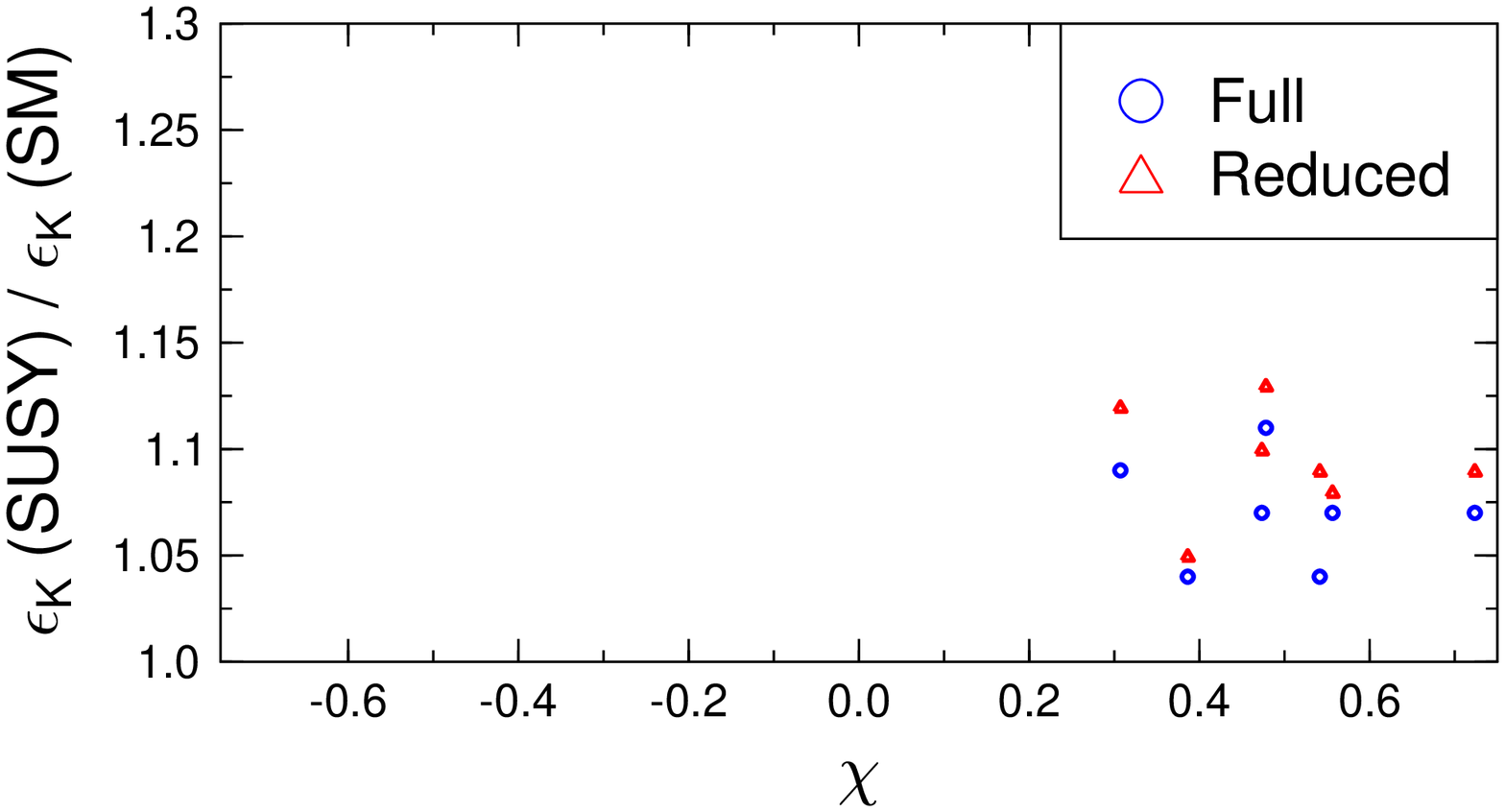}}
\end{tabular}
\parbox[t]{8cm}{\caption{\label{scatter4}\footnotesize 
The ratio of $\epsilon_K$ within the MSSM and the SM in
dependence on $m_{\tilde t_R}$, $m_{\tilde\chi_2}$, $\chi$. In all cases
we decoupled the Higgs boson by setting 
$M_H = 1000\,\mbox{\rm GeV}$ and $\tan\!\beta = 5$. }} 
\end{center}\end{figure}

\begin{figure}\begin{center}
\begin{tabular}{c}
\mbox{\epsfxsize=12.0cm\epsfysize=6.0cm\epsffile{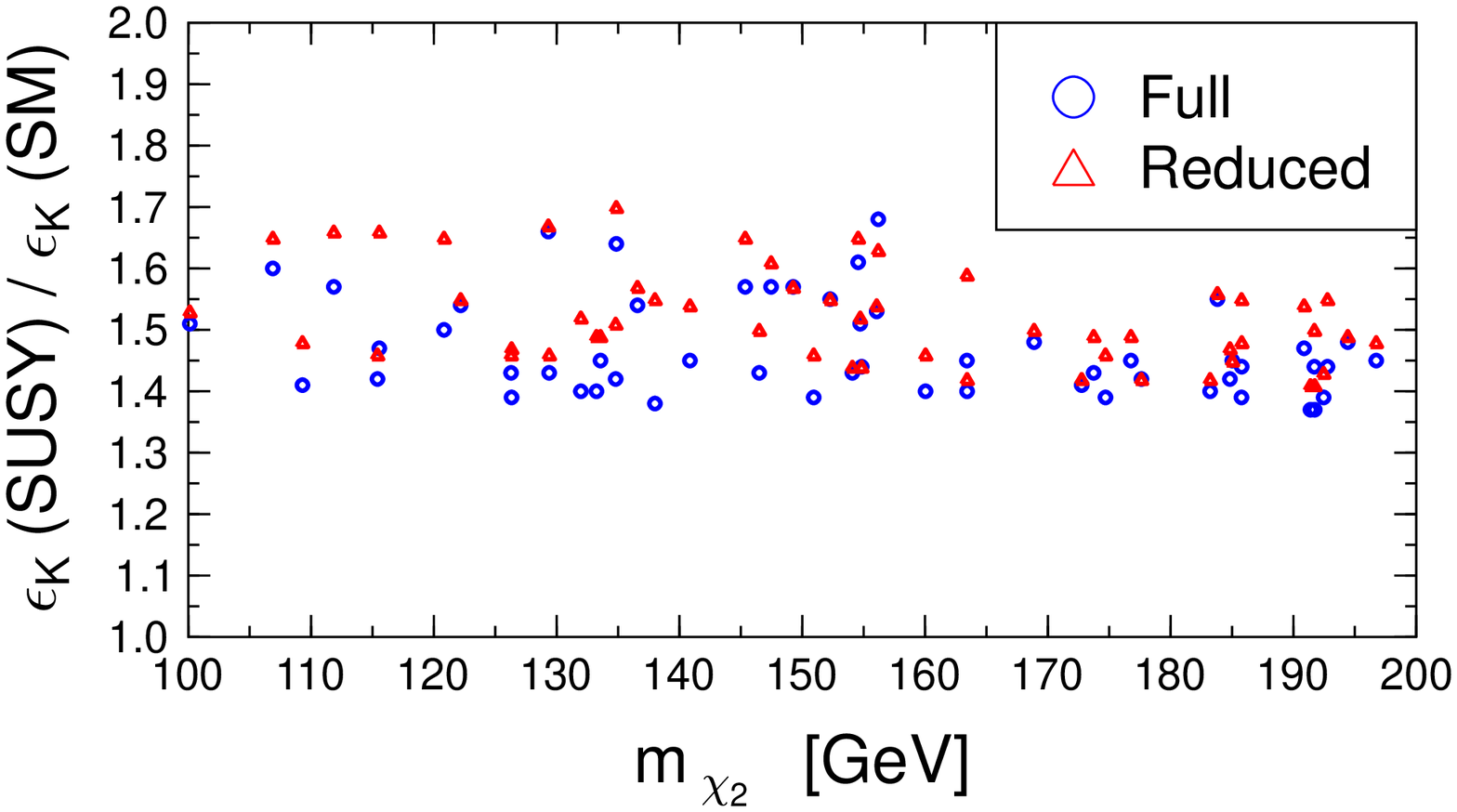}}\\
\mbox{\epsfxsize=12.0cm\epsfysize=6.0cm\epsffile{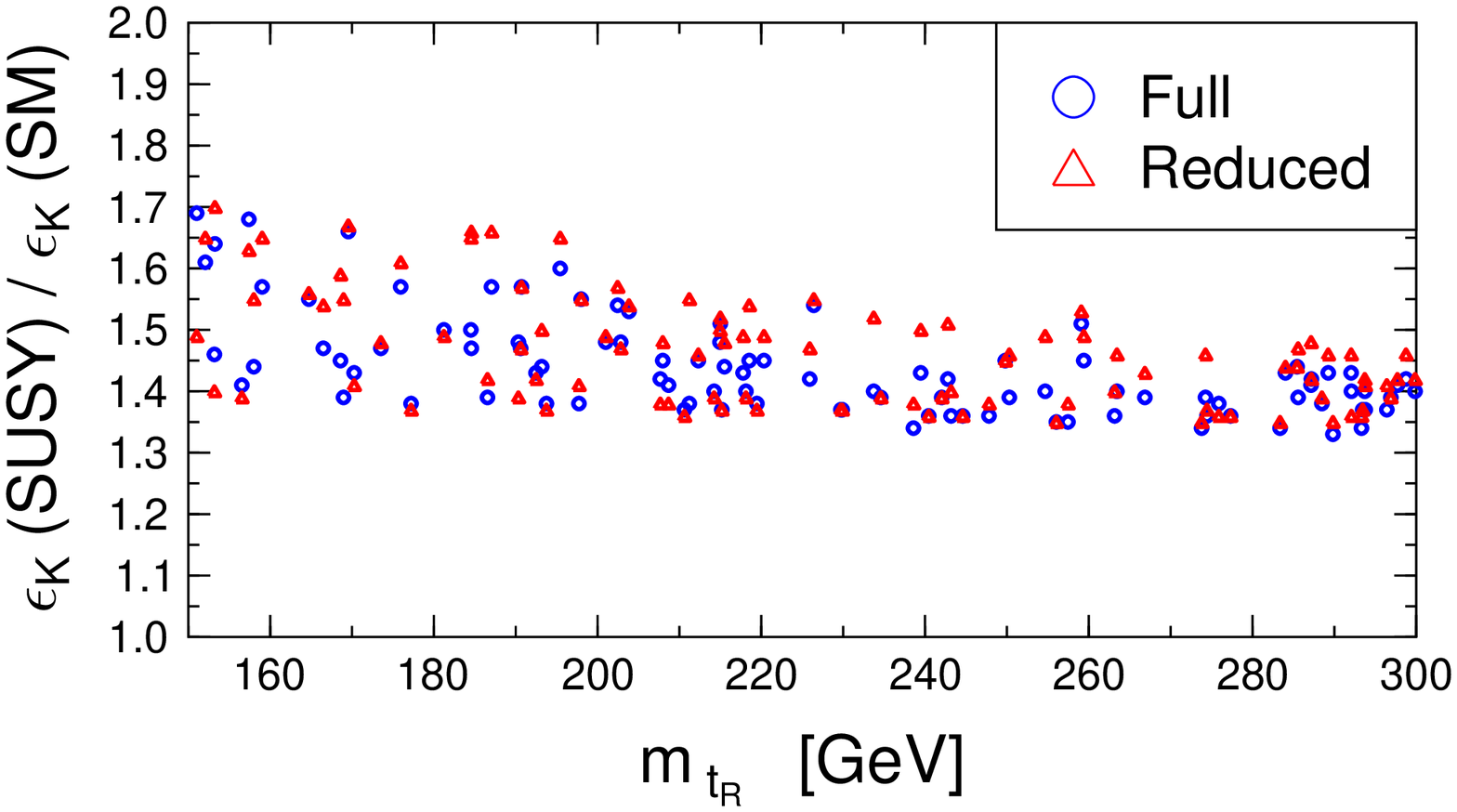}}\\
\mbox{\epsfxsize=12.0cm\epsfysize=6.0cm\epsffile{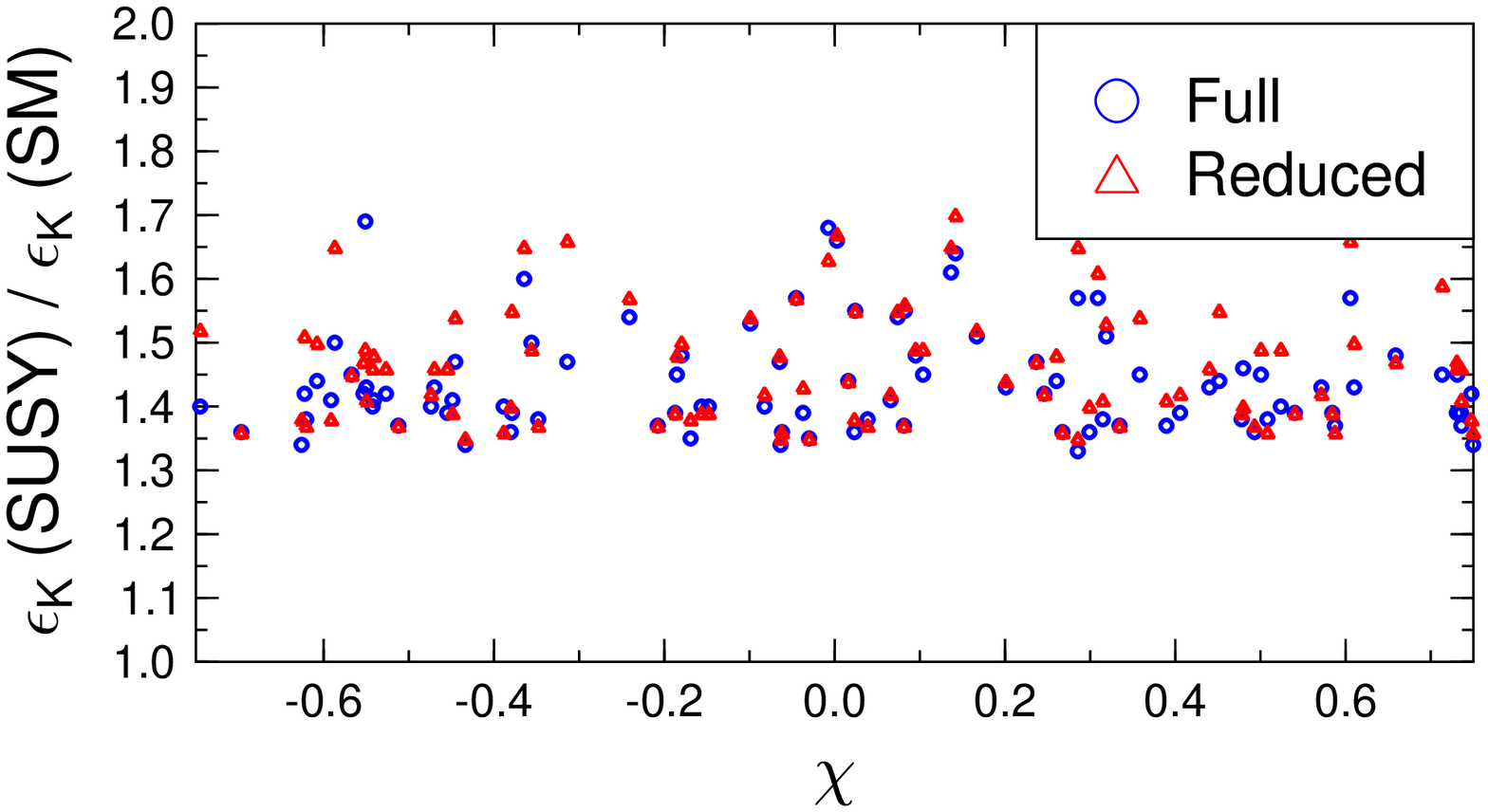}}
\end{tabular}
\parbox[t]{8cm}{\caption{\label{scatter3}\footnotesize 
The ratio of $\epsilon_K$ 
within the MSSM and the SM in
dependence on $m_{\tilde t_R}$, $m_{\tilde\chi_2}$, $\chi$. In all cases
we included the Higgs boson by setting 
$M_H = 100\,\mbox{\rm GeV}$ and $\tan\!\beta = 1.5$. }} 
\end{center}\end{figure}

Obviously -- as can be seen in Figs.\,\ref{scatter1}, \ref{scatter2},
\ref{scatter4}, \ref{scatter3} -- the MSSM might have some considerable
influence on the observables under consideration and it has clearly the
potential to shift the point $(\rho,\eta)$ of the unitarity
triangle in the Wolfenstein parametrization. However, an investigation of
the effect of gluinos and neutralinos on the NLO--level is still missing
and this should be done to have a flavour of the complete picture as
predicted by the MSSM. Furthermore, one should investigate the influence
of gluino--mediated QCD corrections as accomplished for the decay $
b\to s\gamma$ by \cite{Ciu98}.

Nevertheless we have to remark, that the large range of the Bag--parameters
shadows the effect of physics beyond the Standard Model.

\section*{Acknowledgements}

We want to thank G. Buchalla, J. Hewett and J. Wells for pleasant discussions.
Valuable comments of C. Bobeth and J. Urban and the support of DFG, BMBF and 
GSI are gratefully acknowledged. 

\appendix

\section{Inami--Lim Functions}

We consider only the very region of low $\tan\!\beta\approx 1$.
No scalar operator emerges for the effective Hamiltonian.

\subsection{Leading Order Inami--Lim functions}

\bea
S_{ WW}(x_W) =
        \frac{x_W(4 - 11x_W + x_W^2)}
              {4( 1 - x_W )^2} -
         \frac{3 x_W^3\,\log (x_W)}
              {2( 1 - x_W)^3}
\,,\eea

\bea
\lefteqn{\tan^2\!\!\beta\; S_{ HW}(x_W,x_H) =}\nonumber\\&&
  -\frac{x_Hx_W}{4}\,
       \left[\frac{8 - 2x_W}{(1 - x_H)(1 - x_W)} + 
       \frac{(8x_H - 2x_W)\,\log (x_H)}{(1 - x_H)^2(x_H - x_W)} 
       \right. 
\nonumber\\
&&\indone\left.  
-\frac{6\,x_W\,\log (x_W)}{(x_H - x_W)(1 - x_W)^2}\right]
\,,\eea

\bea
\tan^4\!\!\beta\; S_{ HH}(x_W,x_H) =
       \frac{x_Hx_W}{4}\,\frac{\left( 1 - x_H^2 + 2\,x_H\,\log (x_H) \right) }
         {( 1 - x_H )^3}
\,,\eea

\bea
\lefteqn{S_{ \chi_i,\tilde q_a}(x_i,y_a) =}\nonumber\\&&
\frac{x_i + y_a}{(x_i-y_a)^2} - 
\frac{2x_i y_a \log(x_i)}{(x_i-y_a)^3} + 
\frac{2x_i y_a \log(y_a)}{(x_i-y_a)^3}
\,,\eea
\bea
\lefteqn{4S_{ \chi_{i,j},\tilde q_{a,b}}(x_{i,j},y_{a,b}) 
=}\nonumber\\&&
\frac{4x_i^2\log(x_i)}{(x_i-x_j)(x_i-y_a)(x_i-y_b)} - 
\frac{4x_j^2\log(x_j)}{(x_i-x_j)(x_j-y_a)(x_j-y_b)} + 
\nonumber\\[2mm]
&&
\frac{4y_a^2\log(y_a)}{(y_a-x_i)(y_a-x_j)(y_a-y_b)} -
\frac{4y_b^2\log(y_b)}{(y_b-x_i)(y_b-x_j)(y_a-y_b)} 
\,.\eea

\subsection{Next--to--Leading Order functions}

\subsubsection{Standard Model}

\bea
\lefteqn{L_{ WW}^{(1,8)}(x_W)=}\nonumber\\&& 
L_{ WW,tt}^{(1,8)} (x_W) - 
2L_{ WW,tu}^{(1,8)}(x_W) + 
L_{ WW,uu}^{(1,8)}(x_W) 
\nonumber\\[2mm]
&& + 2L_{ W\Phi}^{(1,8)}(x_W) + 
L_{ \Phi\Phi}^{(1,8)}(x_W) \,\,,
\eea

\bea
\lefteqn{L^{(1)}_{ WW,tt}(x_W) =}\nonumber\\&&
   \frac{(4\,x_W + 38\,x_W^2 + 6\,x_W^3)\,\log (x_W)}{(x_W-1)^4}  
-  \frac{3 + 28 x_W + 17 x_W^2}{(x_W-1)^3}
\nonumber \\[2mm]
&& 
+  \frac{(12\,x_W + 48\,x_W^2 + 12\,x_W^3)\,\Litwo(1-1/x_W)}{(x_W-1)^4 } 
\nonumber \\[2mm]
&& 
+  \frac{(24 x_W + 48 x_W^2)\,\Litwo(1-x_W )}{(x_W-1)^4}\,,
\eea

\bea
\lefteqn{2\,L_{ WW,tu}^{(1)}(x_W) =}\nonumber\\&& 
    \frac{2\,(3 + 13 x_W)}{(x_W-1)^2} 
-   \frac{2 x_W\,(5 + 11 x_W)\,\log (x_W)}{(x_W-1)^3 }
\nonumber\\ [2mm]
&& 
-   \frac{12 x_W\,(1 + 3 x_W)\,\Litwo(1-1/x_W)}{(x_W-1)^3}
-   \frac{24 x_W\,(1 + x_W)\Litwo(1-x_W)}{(x_W-1)^3}\,,
\eea

\bea
L^{(1)}_{ WW,uu}(x_W) & =& 3\,\,,
\eea

\bea
\lefteqn{2\,L_{ W\Phi}^{(1)}(x_W)=}\nonumber\\&&
      \frac{4 x_W^2\,(11+13 x_W)}{(x_W-1)^3}
+      \frac{2 x_W^2\,(5 + x_W)(1-9 x_W)\log(x_W)}{(x_W-1)^4}
\nonumber\\[2mm]
&&
-      \frac{24 x_W^2\,(1+4 x_W+x_W^2)\Litwo(1-1/x_W)}{(x_W-1)^4}
\nonumber \\[2mm]
&& 
-      \frac{48 x_W^2\,(1+2 x_W)\Litwo(1-x_W)}{(x_W-1)^4}\,\,,
\eea

\bea
\lefteqn{L_{ \Phi\Phi}^{(1)}(x_W)=}\nonumber\\&&
-     \frac{x_W^2\,(7 + 52 x_W - 11 x_W^2)}{4\,(x_W-1)^3}
+     \frac{3 x_W^3\,(4 + 5 x_W - x_W^2)\log(x_W)}{2\,(x_W-1)^4}
\nonumber\\[2mm]
&&
+     \frac{3 x_W^3\,(3 + 4 x_W - x_W^2)\Litwo(1-1/x_W)}{(x_W-1)^4}
+     \frac{18 x_W^3\,\Litwo(1-x_W)}{(x_W-1)^4}\,\,,
\eea

\bea
\lefteqn{L^{(8)}_{ WW,tt} (x_W) =}\nonumber\\&& 
       \frac{2 x_W (4 - 3 x_W)\,\log (x_W)}{(x_W-1)^3} 
-      \frac{23 -  x_W}{(x_W-1)^2}
\nonumber\\[2mm]
&& 
+      \frac{(8- 12 x_W + 12 x_W^2)\,\Litwo( 1- x_W )}{(x_W-1)^4}
\nonumber\\[2mm]
&& 
-      \frac{(12 x_W -12 x_W^2 - 8 x_W^3 )\,\Litwo(1- 1/x_W)}{(x_W-1)^4}\,,
\eea

\bea
\lefteqn{2 L^{(8)}_{ WW,tu} (x_W) =}\nonumber\\&&  
       \frac{2\,(2 - x_W)\pi^2}{3 x_W} 
-      \frac{(8 - 5 x_W )\log(x_W)}{(x_W-1)^2}
-      \frac{15}{x_W-1}
\nonumber\\ [2mm]
&& 
-  \frac{(6 x_W + 4 x_W^2  )\,\Litwo( 1- 1/x_W)}{x_W(x_W-1)^2}
+  \frac{(8 + 12 x_W - 6 x_W^2)\,\Litwo( 1-  x_W)}{x_W (x_W-1)^2}\,,
\eea

\bea
L^{(8)}_{ WW,uu} (x_W)  &=& -23 + \frac43\,\pi^2\,\,,
\eea

\bea
\lefteqn{2\,L_{ W\Phi}^{(8)}(x_W) =}\nonumber\\&&
   \frac{30 x_W^2}{(x_W-1)^2}
+  \frac{12 x_W^3\log(x_W)}{(x_W-1)^3}
-  \frac{12 x_W^4\,\Litwo(1-1/x_W)}{(x_W-1)^4}
\nonumber\\[2mm]
&& 
-\frac{12 x_W^2\,(2\,- x_W^2)\Litwo(1-x_W)}{(x_W-1)^4}\,\,. 
\eea

\bea
\lefteqn{L_{ \Phi\Phi}^{(8)}(x_W) =}\nonumber\\&&
-    \frac{11 x_W^2\,(1+x_W)}{4\,(x_W-1)^2}
+    \frac{x_W^3\,(4\,-3 x_W)\log(x_W)}{2\,(x_W-1)^3}
\nonumber\\[2mm]
&&
+      \frac{x_W^3\,(3-3 x_W+2 x_W^2)\Litwo(1-1/x_W)}{(x_W-1)^4}
\nonumber\\[2mm]
&&
+   \frac{x_W^2(2+3 x_W-3 x_W^2)\Litwo(1-x_W)}{(x_W-1)^4}\,\,,
\eea

\subsubsection{Two Higgs Doublet Model}

\bea
\lefteqn{L^{(1,8)}_{ \rm THDM}(x_W,x_H) =}\nonumber\\&& 
L_{ WW}(x_W)+\frac{1}{\tan^4(\beta)}\,HH^{(i)}(x_H)
\nonumber\\[2mm]
&&+ \frac{2}{\tan^2(\beta)}\left[L_{ WH}^{(1,8)}(x_W,x_H)\,+
                 L_{ \Phi H}^{(1,8)}(x_W,x_H)\right]\,.
\eea

\begin{eqnarray}\nonumber
\lefteqn{2\,L_{ WH}^{(1)}(x_W,x_H)=}\nonumber\\&&
x_W\left[\frac{2 x_H^2(13+3 x_H)\log(x_H)}{(x_H-1)^3(x_H-x_W)}
\right.
-    \frac{2 x_H\,(9+7 x_H+7 x_W-23 x_W x_H)}{(x_W-1)^2(x_H-1)^2}
\nonumber\\[2mm]
&&\indone
-    \frac{2 x_H^2(18 - 6 x_H\,-44 x_W+13 x_H x_W+9 x_H x_W^2)\log(x_W)}
       {(x_H-1)^2(x_W-1)^3(x_H-x_W)}
\nonumber\\[2mm]
&&\indone
+     \frac{2 x_H x_W(5-27 x_W+6 x_W^2+6 x_H x_W^2)\log(x_W)}
          {(x_H-1)^2(x_W-1)^3(x_H-x_W)}
\nonumber\\[2mm]
&&\indone
-     \frac{24 x_H^2\,\log(x_H)\,\log(x_W)}{(x_H-1)^3(x_H-x_W)}
+      \frac{24 x_H^2\,\Litwo(1-1/x_H)}{(x_H-1)^2(x_H-x_W)}
\nonumber\\[2mm]
&&\indone
\left. 
-       \frac{24 x_H x_W\,(1+x_W)\,\Litwo(1-1/x_W)}{(x_W-1)^3(x_H-x_W)}
-   \frac{48 x_W x_H\,\Litwo(1-x_W)}{(x_W-1)^3(x_H-x_W)}
\right]\,\,, 
\end{eqnarray}

\begin{eqnarray}\nonumber
\lefteqn{2\,L_{ \Phi H}^{(1)}(x_W,x_H) =}\nonumber\\&&
x_W^2\left[\frac{x_H\,(31-15 x_H - 15 x_W-x_H x_W)}
                   {2\,(x_H - 1)^2\,(x_W - 1)^2}\right.
\nonumber\\[2mm]
&&\indone
-           \frac{x_H\,(7+21 x_H - 12 x_H^2)\,\log(x_H)}
              {2\,(x_H - 1)^3\,(x_H - x_W)}
\nonumber\\[2mm]
&&\indone
+          \frac{x_H\,(7 - 9 x_W + 36 x_W^2 - 18 x_W^3)\,\log(x_W)}
          {2(x_H-1)^2(x_H-x_W)(x_W-1)^3}
\nonumber\\[2mm]
&&\indone
+         \frac{x_H^2\,(8 - 36 x_W + 9 x_W^2 + 3 x_W^3)\log(x_W)}
            {(x_H-1)^2(x_H-x_W)(x_W-1)^3}
\nonumber\\[2mm]
&&\indone
-     \frac{x_H^3\,(11 - 45 x_W + 18 x_W^2)\,\log(x_W)}
             {2(x_H-1)^2(x_H-x_W)(x_W-1)^3}
\nonumber\\[2mm]
&&\indone
+       \frac{6 x_H\,\log(x_H)\,\log(x_W)}
              {(x_H-1)^3(x_H-x_W)}
-         \frac{6 x_H(1+x_H-x_H^2)\Litwo(1-1/x_H)}
    {(x_H-1)^2(x_H-x_W)}
\nonumber\\[2mm]
&&\indone
+     \frac{6 x_H\,(1 + 2 x_W^2 - x_W^3)\Litwo(1-1/x_W)}
               {(x_H - x_W)(x_W - 1)^3}
\nonumber\\[2mm]
&&\indone\left.
+   \frac{12 x_H\,\Litwo(1-x_W)}{(x_H - x_W)(x_W - 1)^3}\right]
   \,\,,
\end{eqnarray}     

\bea
L_{ HH}^{(1)}(x_W,x_H) =
     \frac{x_W}{x_H} L_{ \Phi\Phi}^{(1)}(x_H) 
+    6 \log\frac{x_H}{x_W}\,\sum_{i=H,W}\frac{x_i\partial}{\partial x_i}
                 S_{ HH} (x_W,x_H)\,\,,
\end{eqnarray}

\begin{eqnarray}
\lefteqn{2\,L_{ WH}^{(8)} (x_W,x_H) =}\nonumber\\&&
x_W \left[
\frac{24 x_H x_W \Litwo(1 - x_W)}{(x_H - x_W) (x_W-1)^2} \right.
\nonumber\\[2mm]
&&\indone
+\frac{6 x_H^2 (5 x_W - x_H + 3 x_W^2 x_H) \Litwo (1-1/x_W)}
{(x_H-1)^2 (x_H - x_W) (x_W-1)^2 x_W}
\nonumber\\[2mm]
&&\indone
+\frac{6 x_H (2 x_W^2 - 10 x_H x_W + x_H x_W^2) \Litwo (1-1/x_W)}
{(x_H-1)^2 (x_H - x_W) (x_W-1)^2}
\nonumber\\[2mm]
&&\indone
+\frac{6 x_H^2 (5 x_W - x_H -8 x_W^2+ 2 x_H x_W^2) \Litwo (1-x_H)}
{(x_H-1)^2 (x_H - x_W) (x_W-1)^2 x_W}
\nonumber\\[2mm]
&&\indone
+\frac{6 (x_W^2 - x_H x_W + 2 x_H^2 x_W^2) \Litwo(1-x_H)} 
{(x_H-1)^2 (x_H - x_W) (x_W-1)^2}
\nonumber\\[2mm]
&&\indone
+\frac{6 x_H^2 (-x_H+5 x_W) \Litwo(1-1/x_H)}
{(x_H-1)^2 (x_H - x_W) x_W}
\nonumber\\[2mm]
&&\indone
-\frac{6 x_H^2 (5 x_W - x_H -8 x_W^2+ 2 x_H x_W^2) \Litwo (1-x_H/x_W)}
{(x_H-1)^2 (x_H - x_W) (x_W-1)^2 x_W}
\nonumber\\[2mm]
&&\indone
-\frac{6 (x_W^2 - x_H x_W + 2 x_H^2 x_W^2) \Litwo(1-x_H/x_W)} 
{(x_H-1)^2 (x_H - x_W) (x_W-1)^2}
\nonumber\\[2mm]
&&\indone
-\frac{6 x_H (1-x_H - \log (x_H))}{(x_H-1)^2 (x_W-1)}
+\frac{6 x_H (2 x_W-1) \log(x_W)}
{(x_H-1) (x_W-1)^2}
\nonumber\\[2mm]
&&\indone
+\frac{6 x_H^2 (5 x_W - x_H - 8 x_W^2) \log(x_H) \log(x_W)}
{(x_H-1)^2 (x_H - x_W) (x_W-1)^2 x_W}
\nonumber\\[2mm]
&&\indone\left.
+\frac{12 x_H^2 (x_H x_W +x_W^2) \log(x_H) \log(x_W)}
{(x_H-1)^2 (x_H - x_W) (x_W-1)^2}\right]\,,
\end{eqnarray}

\begin{eqnarray}
\nonumber
\lefteqn{2\,L_{ \Phi H}^{(8)}(x_W,x_H)=}\nonumber\\&&
x_W^2\left[\frac{2 x_H + 2 x_W - 11 x_H x_W}
                   {2\,( x_H - 1)\,( x_W - 1) x_W}\right.
\nonumber\\[2mm]
&&\indone
-\frac{(2 x_H^2 - 7 x_H x_W + 2 x_H^2 x_W + 2\,
          x_W^2 + x_H x_W^2)
          \log (x_H)}{2\,( x_H - 1)^2\,(x_H - x_W)\,( x_W - 1) x_W}
\nonumber\\[2mm]
&&\indone
-\frac{x_H\,(7 - 7 x_H + 4 x_W - 6 x_W^2)\,\log (x_W)}
          {2\,( x_H - 1)\,(x_H - x_W)\,( x_W - 1)^2}
\nonumber\\[2mm]
&&\indone
+\frac{(x_H^2 + x_W^2 - 3 x_H^2 x_W^2)\,\log (x_W)}
          {( x_H - 1)\,(x_H - x_W)\,( x_W - 1)^2 x_W}
\nonumber\\[2mm]
&&\indone
-\frac{x_H^2\,(4 - 6 x_W + 3 x_H x_W)\, \log (x_H)\, \log (x_W)}
          {( x_H - 1)^2\,(x_H - x_W)\,( x_W - 1)^2 x_W}
\nonumber\\[2mm]
&&\indone
+\frac{x_H\,(x_H^2 - 3 x_W^2 + 6 x_W^3 - 3 x_W^4)\, \log (x_H)\, \log (x_W)}
          {( x_H - 1)^2\,(x_H - x_W)\,( x_W - 1)^2 x_W^2}
\nonumber\\[2mm]
&&\indone
-\frac{x_H\,(3 x_W^2 + 2 x_H x_W\,(2 + x_W) - x_H^2\,(1 + 2\,
          x_W))\,\Litwo(1-1/x_H)}{( x_H - 1)^2\,(x_H - x_W) x_W^2}
\nonumber\\[2mm]
&&\indone
-\frac{(4 x_H x_W - 6 x_H^2 x_W + 3 x_H^2 x_W^2 - x_W^2)
          \Litwo(1-x_H)}{( x_H - 1)^2\,(x_H - x_W)\,( x_W - 1)^2 x_H}
\nonumber\\[2mm]
&&\indone
-\frac{(4 x_H^2 x_W - 6 x_H^2 x_W^2 - x_H^3 + 3 x_H^3 x_W^2)
          \Litwo(1-x_H)}{( x_H - 1)^2\,(x_H - x_W)\,( x_W - 1)^2 x_W^2}
\nonumber\\[2mm]
&&\indone
+\frac{2 x_H^2\,(6 - x_W^2 - 3 x_H + x_W x_H)\,
          \Litwo(1-1/x_W)}{( x_H - 1)^2\,(x_H - x_W)\,( x_W - 1)^2}
\nonumber\\[2mm]
&&\indone
-\frac{x_H\,(3 x_W^2 + 4 x_H x_W - x_H^2)\,\Litwo(1-1/x_W)}
          {( x_H - 1)^2\,(x_H - x_W)\,( x_W - 1)^2 x_W^2}
\nonumber\\[2mm]
&&\indone
+\frac{(4 x_H x_W - 6 x_H^2 x_W + 3 x_H^2 x_W^2 - x_W^2)
           \Litwo(1-x_H/x_W)}{( x_H - 1)^2\,(x_H - x_W)\,( x_W - 1)^2\,
            x_H}
\nonumber\\[2mm]
&&\indone
+\frac{x_H^2(4 x_W - 6 x_W^2 - x_H + 3 x_H x_W^2)
           \Litwo(1-x_H/x_W)}{( x_H - 1)^2\,(x_H - x_W)\,( x_W - 1)^2\,
            x_W^2}
\nonumber\\[2mm]
&&\indone
\left.-\frac{6 x_H\,\Litwo(1-x_W)}{(x_H-x_W)\,(x_W-1)^2}\right]\,.
\end{eqnarray}

\begin{eqnarray}
L_{ HH}^{(8)}(x_W,x_H) =
      \frac{x_W}{x_H}L_{ \Phi\Phi}^{(8)}(x_H) 
     +  6 \log\frac{x_H}{x_W}\,S_{ HH}(x_W,x_H)\,\,,
\nonumber
\end{eqnarray}

\subsubsection{MSSM, chargino sector}

\bea
\lefteqn{L^{(1)}_{ \chi_i,\tilde q_a}(y_a,x_i) =}\nonumber\\&&
\frac{19y_a^2 + 6y_ax_i + 11x_i^2}
     {(x_i - y_a)^3} - 
\frac{2y_a (2y_a^2 - 11y_ax_i - 9x_i^2)\log(y_a)}
     {(x_i - y_a)^4}  
\nonumber\\[2mm]
&&
-\frac{2x_i (4y_a^2 + 11y_ax_i + 3x_i^2)\log(x_i)}
     {(x_i - y_a)^4} + 
+\frac{2y_ax_i(y_a - 4x_i)}{(x_i - y_a)^4}(\log^2(y_a) - \log^2(x_i))
\nonumber\\[2mm]
&&
+\frac{4y_a(x_i^2 - x_iy_a + 3y_a^2)\Litwo\left(1 - \frac{x_i}{y_a}\right)}
     {(x_i-y_a)^4}
+\frac{4x_iy_a(4x_i-y_a)\Litwo\left(1 - \frac{y_a}{x_i}\right)}
     {(x_i-y_a)^4}
\,,\eea
\bea
\lefteqn{L^{(8)}_{ \chi_i,\tilde q_a}(x_i,y_a) =}\nonumber\\&&
-\frac{7x_i+y_a}{(x_i-y_a)^2} 
-\frac{2x_i(3x_i+4y_a)\log(x_i)}{(x_i-y_a)^3} 
\nonumber\\[2mm]
&&
+\frac{2y_a(4x_i+3y_a)\log(y_a)}{(x_i-y_a)^3} 
-\frac{4y_a(3x_i^2-3x_iy_a-y_a^2)\log(x_i)\log(y_a)}{(x_i-y_a)^4} 
\nonumber\\[2mm]
&&
+\frac{2y_a(6x_i^2-6x_iy_a-y_a^2)}{(x_i-y_a)^4}\log^2(x_i)
-\frac{2y_a^3}{(x_i-y_a)^4}\log^2(y_a)
\nonumber\\[2mm]
&&
+\frac{4(x_i^2+4x_iy_a+y_a^2)}{(x_i-y_a)^3}\Litwo\left(1-\frac{y_a}{x_i}\right)
\,,\eea
\bea
\lefteqn{4L^{(1)}_{ \chi_{i,j},\tilde q_{a,b}}(x_{i,j},y_{a,b}) 
=}\nonumber\\&&
\left[\frac{-24x_i^2}{(x_i-x_j)(x_i-y_a)(x_i-y_b)} \;+\; 
      \{i\leftrightarrow j\}\right]
\nonumber\\[2mm]
&&
-\left[\left(\frac{12y_a[y_a^2(y_b+2x_i+2x_j)-y_ay_b(x_i+x_j)]}
                  {(y_a-y_b)(y_a-x_i)^2(y_a-x_j)}\right.\right.
\nonumber\\[2mm]
&&
\indone\left.\left. +\frac{12y_ax_ix_j(y_b-4y_a)}
            {(y_a-y_b)(y_a-x_i)^2(y_a-x_j)^2}\right)
             \log(y_a)
             \;+\;\{a\leftrightarrow b\}\right]
\nonumber\\[2mm]
&&
-\left[\left(\frac{4x_i[(3y_ay_b+8x_i^2)(y_a+y_b)-5x_iy_ay_b]}
                  {(x_i-x_j)(x_i-y_a)^2(x_i-y_b)^2}\right.\right.
\nonumber\\
&&
\indone\left.\left.-\frac{4x_i[11x_i^3-3x_i(y_a^2+y_b^2)]}
                  {(x_i-x_j)(x_i-y_a)^2(x_i-y_b)^2}\right)
                  \log(x_i) 
      \;+\;\{i\leftrightarrow j\}\right]
\nonumber\\[2mm]
&&
-\left[\frac{4y_a^2[4x_ix_j-3y_a(x_i+x_j)+2y_a^2]}
            {(y_a-y_b)(y_a-x_i)^2(y_a-x_j)^2}
            \log^2(y_a)
        \;+\;\{a\leftrightarrow b\}\right]
\nonumber\\[2mm]
&&
-\left[\frac{4x_i^2[3x_i^2-2x_i(y_a+y_b)+y_ay_b]}
        {(x_i-x_j)(x_i-y_a)^2(x_i-y_b)^2}
       \log^2(x_i)
        \;+\;\{i\leftrightarrow j\}\right]
\nonumber\\[2mm]
&&
-\left[\frac{4y_a(2x_i^2-3x_iy_a+3y_a^2)}
            {(x_i-x_j)(x_i-y_a)^2(x_i-y_b)}
         \Litwo\left(1-\frac{x_i}{y_a}\right)
        \;+\;\{i\leftrightarrow j\,,a\leftrightarrow b\}\right]
\nonumber\\[2mm]
&&
-\left[\frac{4x_i^2(3x_i-y_a)}{(x_i-x_j)(x_i-y_a)^2(x_i-y_b)}
         \Litwo\left(1-\frac{y_a}{x_i}\right)
        \;+\;\{i\leftrightarrow j\,,a\leftrightarrow b\}\right]
\nonumber\\[2mm]
&&
-\left[\left(
      \frac{4y_b^3[3y_b(y_b-y_a-x_i-x_j)+5y_a^2+6x_ix_j]}
            {(x_i-y_a)(x_i-y_b)(x_j-y_a)(x_j-y_b)(y_a-y_b)^2}\right.\right.
\nonumber\\[2mm]
&&
\indone\left.\left.-\frac{4y_b[y_a^2y_b(y_a+x_i+x_j)+2x_ix_jy_a(2y_b-y_a)]}
            {(x_i-y_a)(x_i-y_b)(x_j-y_a)(x_j-y_b)(y_a-y_b)^2}\right)
         \Litwo\left(1-\frac{y_a}{y_b}\right)\right.
\nonumber\\[2mm]
&&
\indthree\left.+\vphantom{\frac{|}{|}}
        \;\{a\leftrightarrow b\}\right]
\,,\eea
\bea
4\lefteqn{L^{(8)}_{ \chi_{i,j},\tilde q_{a,b}}(x_{i,j},y_{a,b}) 
=}\nonumber\\&&
\left[\frac{4x_i(2x_i+x_j)}{(x_i-x_j)(x_i-y_a)(x_i-y_b)}\log(x_i)
      \;+\;\{i\leftrightarrow j\}\right]
\nonumber\\[2mm]
&&
+\left[\frac{4y_a(y_a+x_i+x_j)}{(y_a-y_b)(y_a-x_i)(y_a-x_j)}\log(y_a)
      \;+\;\{a\leftrightarrow b\}\right]
\nonumber\\[2mm]
&&
-\left[\frac{12x_i^2}{(x_i-x_j)(x_i-y_a)(x_i-y_b)}\log^2(x_i)
      \;+\;\{i\leftrightarrow j\}\right]
\nonumber\\[2mm]
&&
-\left[\frac{12y_a^2}{(y_a-y_b)(y_a-x_i)(y_a-x_j)}\log^2(y_a)
      \;+\;\{a\leftrightarrow b\}\right]
\nonumber\\[2mm]
&&
-\left[\frac{4y_a^2[3(x_i^2-y_a^2)-6x_iy_b+5y_ay_b]}
              {(x_i-y_a)^2(x_i-y_b)(x_j-y_a)(y_a-y_b)}
       \Litwo\left(1-\frac{x_i}{y_a}\right)\right.
\nonumber\\[2mm]
&&
\indthree\left.+\vphantom{\frac{|}{|}}
      \;\{i\leftrightarrow j\,,a\leftrightarrow b\}\right]
\nonumber\\[2mm]
&&
-\left[\left(\frac{4x_i^2[x_i^2(2x_i-2x_j-5y_a+y_b)+3y_a^2(x_i-x_j)]}
         {(x_i-x_j)(x_i-y_a)^2(x_i-y_b)(x_j-y_a)(y_a-y_b)}\right.\right.
\nonumber\\[2mm]
&&
\indone\left.\left.+\frac{4x_i^3[x_j(5y_a+y_b)-2y_ay_b]}
         {(x_i-x_j)(x_i-y_a)^2(x_i-y_b)(x_j-y_a)(y_a-y_b)}\right)
       \Litwo\left(1-\frac{y_a}{x_i}\right)\right.
\nonumber\\[2mm]
&&
\indthree\left.+\vphantom{\frac{|}{|}}
        \;\{i\leftrightarrow j\,,a\leftrightarrow b\}\right]
\nonumber\\[2mm]
&&
-\left[\left(\frac{4y_a^2[6y_ay_b-3(x_i+x_j)(y_a+y_b)]}
                 {(y_a-y_b)(x_i-y_a)(x_i-y_b)(x_j-y_a)(x_j-y_b)}\right.\right.
\nonumber\\[2mm]
&&
\indone\left.\left.+\frac{4y_a^2[2x_ix_j(2y_a-3y_b)+y_ay_b(x_i+x_j)]}
                 {(y_a-y_b)(x_i-y_a)(x_i-y_b)(x_j-y_a)(x_j-y_b)}\right)
       \Litwo\left(1-\frac{y_b}{y_a}\right)\right.
\nonumber\\[2mm]
&&
\indthree\left.+\vphantom{\frac{|}{|}}
        \;\{a\leftrightarrow b\}\right]
\nonumber\\[2mm]
&&
-\left[\frac{4x_i^2[2x_i^2-3x_j^2-6(y_ay_b+x_ix_j)+6(x_i+x_j)(y_a+y_b)]}
            {(x_i-x_j)(x_i-y_a)(x_i-y_b)(x_j-y_a)(x_j-y_b)}
       \Litwo\left(1-\frac{x_j}{x_i}\right)\right.
\nonumber\\[2mm]
&&
\indthree\left.+\vphantom{\frac{|}{|}}
        \;\{i\leftrightarrow j\}\right]
\,.\eea

\newpage

\begin{table}
\,\mbox{\epsfxsize=14cm\epsffile{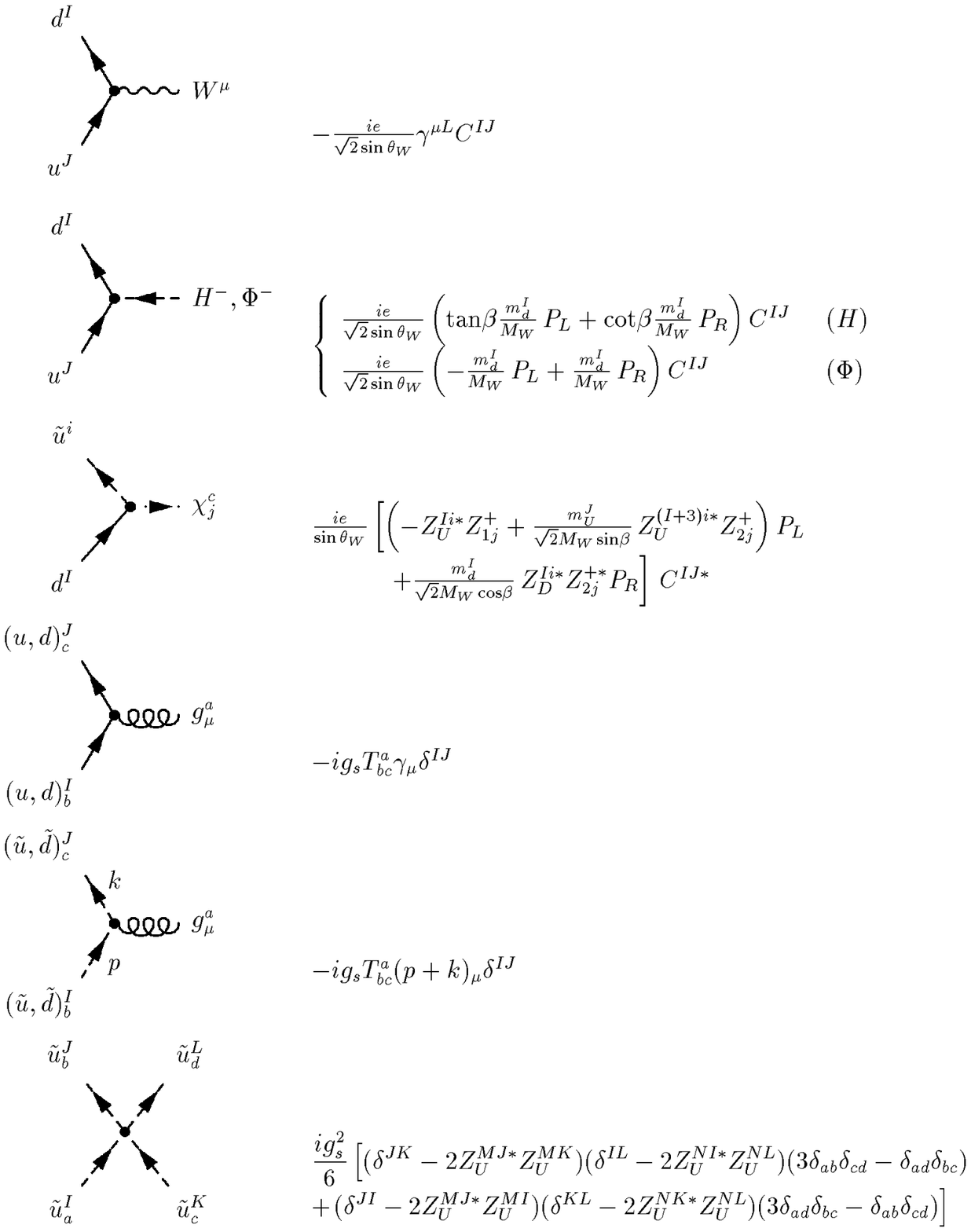}}
\parbox[t]{13.5cm}{\caption{\label{Feyns}
\footnotesize The Feynman--rules employed throughout this article.
Note, that only the part proportional to the strong coupling
constant enters the four--squark vertex.}}
\end{table}


\newpage

\begin{thebibliography}{99}

\bibitem{Hab85} H. E. Haber, G. L. Kane; Phys. Rep. 117 (1985) 75.
\bibitem{Gun86} J. F. Gunion, H. E. Haber; Nucl. Phys. B272 (1986) 1.
\bibitem{WB92} J. Wess, J. Bagger; \textsl{Supersymmetry and Supergravity},
               Princeton University Press, Princeton, New Jersey 
               (1992).
\bibitem{Hab93} H. E. Haber; \textsl{Introductory Low-Energy Supersymmetry},
                Proceedings of the 1992 Theoretical Advanced
                Study Institute in Particle Physics, World Scientific,
                Singapore (1993) 589.
\bibitem{Mai80} L. Maiani, Proc. Summer School of Gif-sur-Yvette, 
                Paris 1980, p. 3;
\bibitem{Vel81} M. Veltman, Acta Phys. Pol. B12 (1981) 437;
\bibitem{Wit81} E. Witten, NPB 188 (1981) 513.
\bibitem{Dre96} M. Drees; \textsl{An Introduction To Supersymmetry},
                Lectures given at Inauguration Conference of the Asia Pacific 
                Center for Theoretical Physics (APCTP), Seoul, Korea (1996); 
                hep-ph/9611409. 
\bibitem{Daw97} S. Dawson; \textsl{The MSSM and why it works}, Lectures given 
                at the 1997 TASI summer school, "Supersymmetry, Supergravity, 
                and Supercolliders" (1997); hep-ph/9712464.
\bibitem{Ber91} S. Bertolini, F. Borzumati, A. Masiero, G. Ridolfi, 
                Nucl. Phys. B353 (1991) 591;
\bibitem{Bar93} R. Barbieri, G. F. Giudice, Phys. Lett. B309 (1993) 86;
\bibitem{Osh93} N. Oshimo, Nucl. Phys. B404 (1993) 20;
\bibitem{Gar93} R. Garisto, J. N. Ng, Phys. Lett. B315 (1993) 372;
\bibitem{Dia93} M. A. Diaz, Phys. Lett. B304 (1993) 278;
\bibitem{Oka93} Y. Okada, Phys. Lett. B315 (1993) 119;
\bibitem{Bor94} F. Borzumati, Z. Phys. C63 (1994) 291;
\bibitem{Nat94} P. Nath, R. Arnowitt, Phys. Lett. B336 (1994) 395;
\bibitem{Ber95} S. Bertolini, F. Vissani, Z. Phys. C67 (1995) 513;
\bibitem{Lop95} J. Lopez et al., Phys. Rev. D51 (1995) 147;
\bibitem{Bra94} G. C. Branco, G. C. Cho, Y. Kizukuri, N. Oshimo, 
                Phys. Lett. B337 (1994) 316; 
\bibitem{Nat94a} P. Nath, R. Arnowitt, Phys. Lett. B336 (1994) 395;
\bibitem{Bra96} G. C. Branco, W. Grimus, L. Lavoura, 
                Phys. Lett. B380 (1996) 119;
\bibitem{Bri96} A. Brignole, F. Feruglio, F. Zwirner, Z. Phys. C71 (1996) 679.
\bibitem{Hew96} J. L. Hewett, James D. Wells;
                Phys. Rev. D55 (1997) 5549.
\bibitem{Hew96a} J. L. Hewett, T. Takeuchi, S. Thomas;
             \textsl{Indirect Probes of New Physics},
             to appear in {\it Electroweak Symmetry Breaking and Beyond the 
             Standard Model}, ed. T. Barklow, S. Dawson, H. Haber, S. Siegrist,
             World Scientific
             hep-ph/9603391.
\bibitem{Hew98} J. L. Hewett; hep-ph/9803370.
\bibitem{Bur90} A. J. Buras, M. Jamin, P. H. Weisz; 
                Nucl. Phys. B347 (1990) 491.
\bibitem{Her95} S. Herrlich, U. Nierste; Phys. Rev. D52 (1995) 6505.
\bibitem{Buc96} G. Buchalla, A. J. Buras, M. E. Lautenbacher;
                Rev. Mod. Phys. 68 (1996) 1125.
\bibitem{Urb98} J. Urban, F. Krauss, U. Jentschura, G. Soff;
                Nucl. Phys. B523 (1998) 40.
\bibitem{Ciu98} M. Ciuchini, G. Degrassi, P. Gambino, G. F. Giudice;
                Nucl. Phys. B534 (1998) 3.
\bibitem{Gab96}  F. Gabbiani, E. Gabrielli, A. Masiero, L. Silvestrini;
                 Nucl. Phys. B477 (1996) 321.
\bibitem{Mis97} M. Misiak, S. Pokorski, J. Rosiek; \textsl{Supersymmetry
                and FCNC effects}, to appear in the Review Volume ``Heavy 
                Flavours II'', eds. A. J. Buras and M. Lindner, Advanced 
                Series on Directions in High Energy Physics, World Scientific 
                Publishing Co., Singapore, hep-ph/9703442
\bibitem{Mas97} A. Masiero, L. Silvestrini; \textsl{Two lectures on FCNC and 
                CP violation in Supersymmetry}, Lectures given at 
                International School of Subnuclear Physics, 35th Course: 
                Highlights: 50 Years Later, Erice, Italy, 26 Aug - 4 Sep 1997,
                and given at International School of Physics, 'Enrico Fermi': 
                Heavy Flavor Physics, hep-ph/9711401;
\bibitem{Bag97} Jonathan A. Bagger, Konstantin T. Matchev, Ren-Jie Zhang;
                Phys. Lett. B412 (1997) 77.
\bibitem{Cha82} A. H. Chamseddine, R. Arnowitt, Pran Nath;
                Phys. Rev.Lett. 49 (1982) 970.
\bibitem{Bar82} R. Barbieri, S. Ferrara, C. A. Savoy;
                Phys. Lett. 119B (1982) 343.
\bibitem{Hal83} L. Hall, J. Lykken, S. Weinberg;
                Phys. Rev. D27 (1983) 2359.
\bibitem{Nat83} Pran Nath, R. Arnowitt, A. H. Chamseddine;
                Nucl. Phys. B227 (1983) 121.
\bibitem{GNO95} T. Goto, T. Nihei, Y. Okada; 
                Phys. Rev. D53 (1996) 5233; Erratum-ibid. D54 (1996) 5904.
\bibitem{Boe97} W. de Boer;
                Acta Phys. Polon. B28 (1997) 1395.
\bibitem{Boe97a}  W. de Boer, R. Ehret, A. V. Gladyshev, D. I. Kazakov;
                 hep-ph/9712376.
\bibitem{Erl98} J. Erler, D. M. Pierce;
                hep-ph/9801238.
\bibitem{Ros90} J. Rosiek; Phys. Rev. D41 (1990) 3464. 
                           Err. in hep-ph/9511250.
\bibitem{IL81} T. Inami, C. N. Lim; Prog. Theor. Phys. 65 (1981) 297.
\bibitem{GW80} F. J. Gilman, M. B. Wise;
               Phys. Lett. 93B (1980) 129.
\bibitem{Bur90a} A. J. Buras, P. H. Weisz; Nucl. Phys. B333 (1990) 66.
\bibitem{Bar78} W. A. Bardeen, A. J. Buras, D. W. Duke, T. Muta;
                Phys. Rev. D18 (1978) 3998.
\bibitem{mathem} S. Wolfram; \textsl{Mathematica: A System for Doing 
                 Mathematics by Computer}, Addison-Wesley, 
                 Reading (1993).
\bibitem{FArts} J. K{\"u}blbeck, M. B{\"o}hm, A. Denner;
                Comp. Phys. Comm. 60 (1990) 165  
\bibitem{PDG} Particle Data Group;
              Phys. Rev. D54 (1996) 1. 

%
%
%
%
\end{thebibliography}
\end{document}